\documentclass[twocolumn,superscriptaddress,aps,prb,showpacs,showkeys,amsmath,preprintnumbers]{revtex4}

\usepackage{graphicx}												
\usepackage{color}
\usepackage{siunitx}	
\usepackage{scrtime}												
\usepackage[T1]{fontenc}
\usepackage[latin1]{inputenc}

\newcommand{\SdH}{Shubnikov-de Haas}
\newcommand{\YRS}{YbRh$_2$Si$_2$}													
\newcommand{\LRS}{LuRh$_2$Si$_2$}											
\newcommand{\mK}{\milli\kelvin}
\newcommand{\kT}{\kilo\tesla}
\newcommand{\mT}{\milli\tesla}
\newcommand{\Freq}[1]{\SI{#1}{\kT}}

\newcommand{\mstar}{\ensuremath{m^{\star}}}
\newcommand{\EF}{\ensuremath{E_{\text F}}}
\newcommand{\zSi}{\ensuremath{z_{\text Si}}}
\newcommand{\Jp}{\ensuremath{\text J3^{\prime}}}
\newcommand{\me}{\ensuremath{m_{\text e}}}

\begin{document}

\preprint{Manuscript: version 2 Date: \today\  \thistime}

\title{Electronic Structure of \LRS:  ''Small`` Fermi Surface Reference to \YRS.}

\author{Sven Friedemann}
	\email{sf425@cam.ac.uk}

\author{Swee K Goh}

\author{Patrick M C Rourke \footnote{Current address:
National Research Council Canada, 1200 Montreal Road, Ottawa, Ontario,
K1A 0R6, Canada.}}

\author{Pascal Reiss}

\author{Michael L Sutherland}

\author{F Malte Grosche}

\affiliation{Cavendish Laboratory, University of Cambridge, JJ Thomson Avenue, CB3 0HE Cambridge, UK}

\author{Gertrud Zwicknagl}

\affiliation{Institute for Mathematical Physics, TU Braunschweig, Mendelssohnstraße 3, 38106 Braunschweig, Germany}

\author{Zachary Fisk}

\affiliation{Department of Physics and Astronomy, University of California, Irvine, CA 92697-4575, USA}

\date{\today }

\begin{abstract}
We present band structure calculations and quantum oscillation measurements on \LRS, which is an ideal reference to the intensively studied quantum critical heavy-fermion system \YRS.
Our band structure calculations show a strong sensitivity of the Fermi surface on the position of the silicon atoms $\zSi$ within the unit cell. Single crystal structure refinement and comparison of predicted and observed quantum oscillation frequencies and masses yield $\zSi = \SI{0.379}{c}$ in good agreement with numerical lattice relaxation. 
 This value of \zSi\ is suggested for future band structure calculations on \LRS\ and \YRS.
\LRS\ with a full $f$ electron shell represents the ``small'' Fermi surface configuration of \YRS. Our experimentally and \textit{ab intio} derived quantum oscillation frequencies of \LRS\ show strong differences with earlier measurements on \YRS. 
Consequently, our results confirm the contribution of the $f$ electrons to the Fermi surface of \YRS\ at high magnetic fields.
%
Yet the limited agreement with refined fully itinerant local density approximation calculations  highlights the need for more elaborated models to describe the Fermi surface of \YRS.
\end{abstract}
%
\maketitle

%
%
\section{Introduction}
\LRS\ is an intermetallic compound crystallising in the tetragonal ThCr$_2$Si$_2$ structure. It is isostructural to the heavy fermion material \YRS\ with almost identical lattice parameters (cf.\ Tab~\ref{tab:latpar}). Moreover, \LRS\ has the same electronic configuration as \YRS\ except that \LRS\ has a completely filled 4$f$ electron shell whereas in \YRS\ the trivalent configuration of Yb has one electron missing in the 4$f$ shell. This missing electron can be regarded as a 4$f$ hole. In analogy to the Ce based heavy fermion systems with one electron in the 4$f$ shell, the hole in the 4$f$ shell of \YRS\ is the basis for the rich physics of this system. The scattering of conduction electrons from the 4$f$ hole---known as the Kondo effect---gives rise to new states near the Fermi energy which can be regarded as composite quasiparticles formed of the $f$ hole and the conduction electrons. These quasiparticles carry the same quantum numbers as non-interacting electrons, however, they posses  highly renormalised properties like a hugely enhanced mass. At temperatures below $T_{\text N} = \SI{70}{\mK}$, \YRS\ undergoes a transition into an antiferromagnetically ordered state which can be fully suppressed with a small critical field of \SI{60}{\mT} \cite{Gegenwart2002}. At zero temperature the transition from this magnetically ordered state to the paramagnetic state represents a quantum critical point (QCP). \YRS\ has emerged as a prototypical system for a new class of QCPs which need descriptions beyond the order parameter notion \cite{Si2001}. Hall effect measurements show a crossover in the Hall coefficient which sharpens to a jump at the QCP in the extrapolation to zero temperature \cite{Paschen2004,Friedemann2010b}. This points towards a sudden reconstruction of the Fermi surface which is not expected at a QCP where the magnetic order parameter evolves continuously. Rather, these results suggest the breakdown of the Kondo effect and the disintegration of the composite quasiparticles. Within this scenario, the Fermi surface evolves from ''large``, including the $f$ electron states in the paramagnetic phase, to ''small`` in the magnetic phase formed of the non-$f$ states only. The latter configuration is paralleled by \LRS\ as here the completely filled $f$ states lie well below the Fermi energy and do not contribute to the Fermi surface. In fact, when tuning across the QCP towards the suggested ``small'' Fermi surface configuration, the Hall coefficient crossover in \YRS\ has a trend towards the Hall coefficient of \LRS\ \cite{Friedemann2010c}.

\begin{table}
  \begin{tabular}{llll}
    \hline
    Compound & $a, b$ [\si{\angstrom}] & $c$ [\si{\angstrom}] & \zSi\ [$c$] \\
    \hline
    \LRS & $4.006(1)$ & $9.838(3)$ & $0.379(2)$ \\
    \YRS & $4.007(1)$ & $9.858-9.862$ & $0.379(2)$ \\
    \hline
  \end{tabular}
   \caption{Lattice parameters of \LRS\ were obtained at room temperature from x-ray diffraction measurements of powdered single crystals \cite{Caroca-Canales2010}. The Wyckoff  positions \zSi\ were deduced from single crystalline structure refinement \cite{Cardoso2011}. For \YRS\ the height of the unit cell, $c$, has been reported to depend on the exact Rh content \cite{Wirth2012} whereas no change in the \zSi\ parameter was resolved \cite{Cardoso2011}.}
  \label{tab:latpar}
\end{table}

Moreover, at a magnetic field $\mu_0H_0 \approx \SI{10}{\tesla}$ a second transition is observed in transport, thermodynamic, and quantum oscillation measurements on \YRS\ \cite{Pfau2013,Rourke2008}. With the Kondo temperature and $H_0$ representing similar energy scales and exhibiting scalable pressure dependencies, one might associate $H_0$ with the polarisation of the Kondo singlet states and a suppression of the Kondo effect, yielding again a ''small`` Fermi surface above $H_0$ [\onlinecite{Gegenwart2006}]. However, the continuous evolution of the quantum oscillation frequencies at $H_0$ rather indicates a complete depopulation of one Fermi surface branch. 
This field dependence is a strong indicator for the ``large'' Fermi surface character as this is not expected for a system with localized $f$ electrons. In fact, renormalised band structure calculations suggest a non-linear dependencies of the Fermi surface cross sections in magnetic field \cite{Zwicknagl2011} which lead to changing frequencies in quantum oscillation measurements.
However, the comparison of measured angular dependences in \YRS\ with local density approximation (LDA) calculations of both \LRS\ and \YRS\ representing the ``small'' and ``large'' Fermi surface, respectively, were inconclusive \cite{Rourke2008,Sutton2010}. While LDA is known to fall short of modelling Yb-based heavy fermion systems \cite{Herbst1984} LDA has proved reliable for normal metals as \LRS\ including the ``small'' Fermi surface configuration. Consequently, the discrepancies between frequencies measured on \YRS\ and the calculated ``small'' Fermi surface point towards the $f$ electron contributing while the discrepancies with the calculated ``large'' Fermi surface are expected. 
All earlier calculations \cite{Knebel2006,Rourke2008,Sutton2010} were based on generic lattice parameters different from the refined values obtained in the present study.

For many systems of the ThCr$_2$Si$_2$ structure there is a strong sensitivity of the electronic structure to the precise crystal structure. In particular, the Wyckoff parameter specifying the position  \zSi\ of the Si atoms has a large influence on the band structure as demonstrated for CeRu$_2$Si$_2$ and LaRu$_2$Si$_2$ \cite{Suzuki2010}. For LaRu$_2$Si$_2$ a lattice relaxation yielded a value \zSi\ far from the experimental value and, moreover, led to a Fermi surface topology in much better agreement with quantum oscillation measurements. In the iron pnictides, the Fe-As-Fe angle which is an equivalent measure of the same Wyckoff parameter  \zSi\ is crucial for the nesting of the Fermi surface sheets and for the optimum superconducting transition temperature \cite{Kimber2009,Lee2008}.   Previous electronic band structure calculations on \LRS\ and \YRS\ used either a generic value $\zSi= \SI{0.375}{c}$ or did not reveal the value used \cite{Friedemann2010c,Wigger2007,Knebel2006,Jeong2006,Rourke2008,Sutton2010}. For \LRS, also, a strong dependence of the Fermi surface topology with respect to \zSi\ has been found in a comprehensive band structure calculation \cite{Reiss2013}.

Here, we present a detailed study of the electronic structure of \LRS\ using \SdH\ measurements which we compare with band structure calculations. 
Best agreement between predicted and observed quantum oscillation frequencies is obtained at the precisely determined experimental value  $\zSi = 0.379 c$. 
Future electronic structure calculations should use the experimental Wyckoff parameter \zSi\ and precisely determined lattice parameters.
Our electronic structure investigations on \LRS\ provide a more accurate reference for the ``small'' Fermi surface configuration. In fact, we find significant modifications arising from the corrected silicon position \zSi. Nevertheless, we find even stronger differences between the experimental results on \YRS\ and our refined ``small'' Fermi surface calculations. 

In addition, we find indications for some of the quantum oscillations in \YRS\ with frequencies between \Freq{4} and \Freq{7} to arise from harmonics. If this proves to be right, band structure calculations may be compared to the remaining frequencies below \Freq{4} and above \Freq{7}, only. We suggest further experiments to check for these potential harmonics.


%
%

\section{Band structure calculations}
\label{sec:BS}
%
%
\subsection{Computational Details}
\label{subsec:BS_Method}
We use the WIEN2k density functional theory code to perform band structure calculations \cite{Blaha2011}. The band structure and the Fermi surface topology are found to be very sensitive to details of the crystal structure, particularly the position of the Si atoms \zSi \cite{Reiss2013}. Therefore, very accurate powder x-ray diffraction measurements and single crystal structure refinement were performed on a piece of the same single crystal used for the \SdH-measurements. The obtained crystallographic parameters of the tetragonal unit cell with space group $I4/mmm$ (\# 139) as given in Tab.~\ref{tab:latpar} with
the relative atomic positions Lu (0,0,0), Rh (0,1/2,1/4), and Si (0,0,\zSi) 
were used for the band structure calculations. Band energies were calculated on a \num{800000} $k$-point mesh in the Brillouin zone using the Perdew-Burke-Ernzerhof generalized gradient approximation to the exchange-correlation potential \cite{Perdew1996}, $R K_{\text{max}} = 7$, and an energy range \SIrange{-6}{5}{Ry}. This corresponds to a valence band treatment of 5$s$, 5$p$, 5$d$, 4$f$, and 6$s$ electrons for Lu, 4$p$, 4$d$, and 5$s$ for Rh, and 3$s$ and 3$p$ for Si. WIEN2k is based on a full potential augmented plane wave and local orbits approach \cite{Blaha2011}. Relativistic effects and spin-orbit (SOC) coupling are included on a one-electron level with relativistic local orbits used for Lu and Rh whereas Si is well approximated by non-relativistic local orbitals due to its light mass. 
The density of states (DOS) was calculated utilizing the tetrahedron method.
Fermi surfaces are visualized with the XCrysDen program \cite{Kokalj1999}. Extremal orbits and effective masses  were calculated using the SKEAF algorithm  \cite{Rourke2012} at an interpolation of 350 in the full Brillioun zone which corresponds to an interpolation factor of $\approx 5$ with respect to the $k$-mesh of the band structure calculations. This allows to accurately detect frequencies down to \Freq{0.017}.

%
%
\subsection{Electronic structure and Fermi surface}
\label{subsec:BS_Results}

\begin{figure}%
\includegraphics[width=.9\columnwidth]{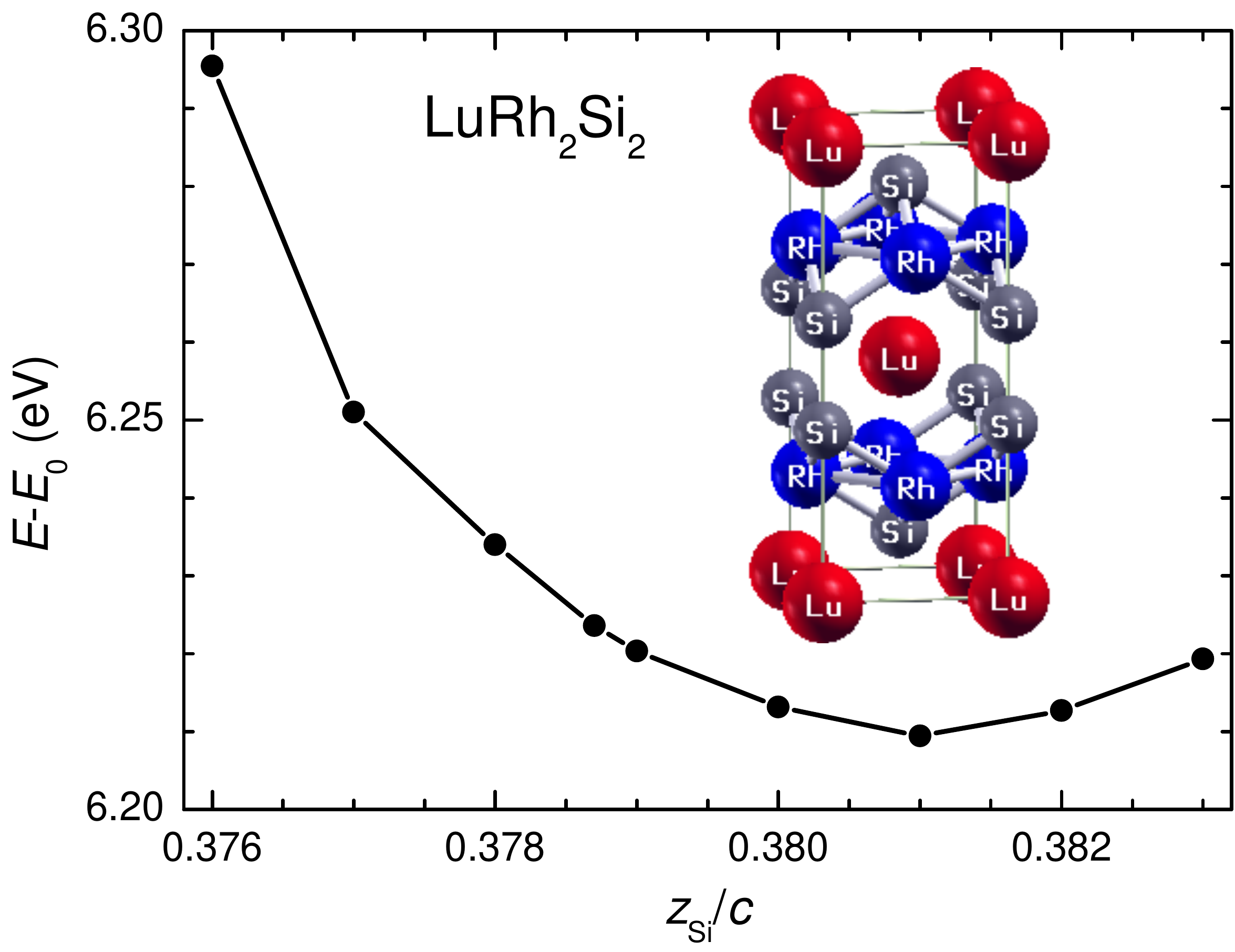}
\caption{Total energy of \LRS\ in our bandstructure calculations as a function of \zSi. A constant offset of $E_0 = -673$keV has been subtracted. The inset depicts the crystal structure.}%
\label{fig:totEn}%
\end{figure}

We start by optimizing the crystal structure with respect to the Si position \zSi\ by minimizing the total energy obtained from our band structure calculations whilst keeping the lattice parameters $a$ and $c$ fixed at their precisely determined experimental values (cf.\ Tab.~\ref{tab:latpar}). 
As can be seen from Fig.~\ref{fig:totEn} the total energy is minimal for $\zSi = \SI{0.381}{c}$. 
This value agrees with that obtained from  crystal structure refinement $\zSi = \SI{0.379(2)}{c}$ within experimental accuracy \cite{Cardoso2011}. Moreover, the energy difference between  the experimental and relaxed Si position amounts to $\approx \SI{10}{\milli\electronvolt}$ ($\approx \SI{100}{\kelvin}$) only. 
Indeed as we shall see from the comparison of calculated orbits with experimentally observed quantum oscillation frequencies, we find best agreement for the experimental \zSi\ (cf.\  Ref.~\cite{Reiss2013}). In the following we present results obtained for the experimentally determined value $\zSi=\SI{0.379}{c}$, if not stated otherwise.

\begin{figure}%
\includegraphics[width=.8\columnwidth]{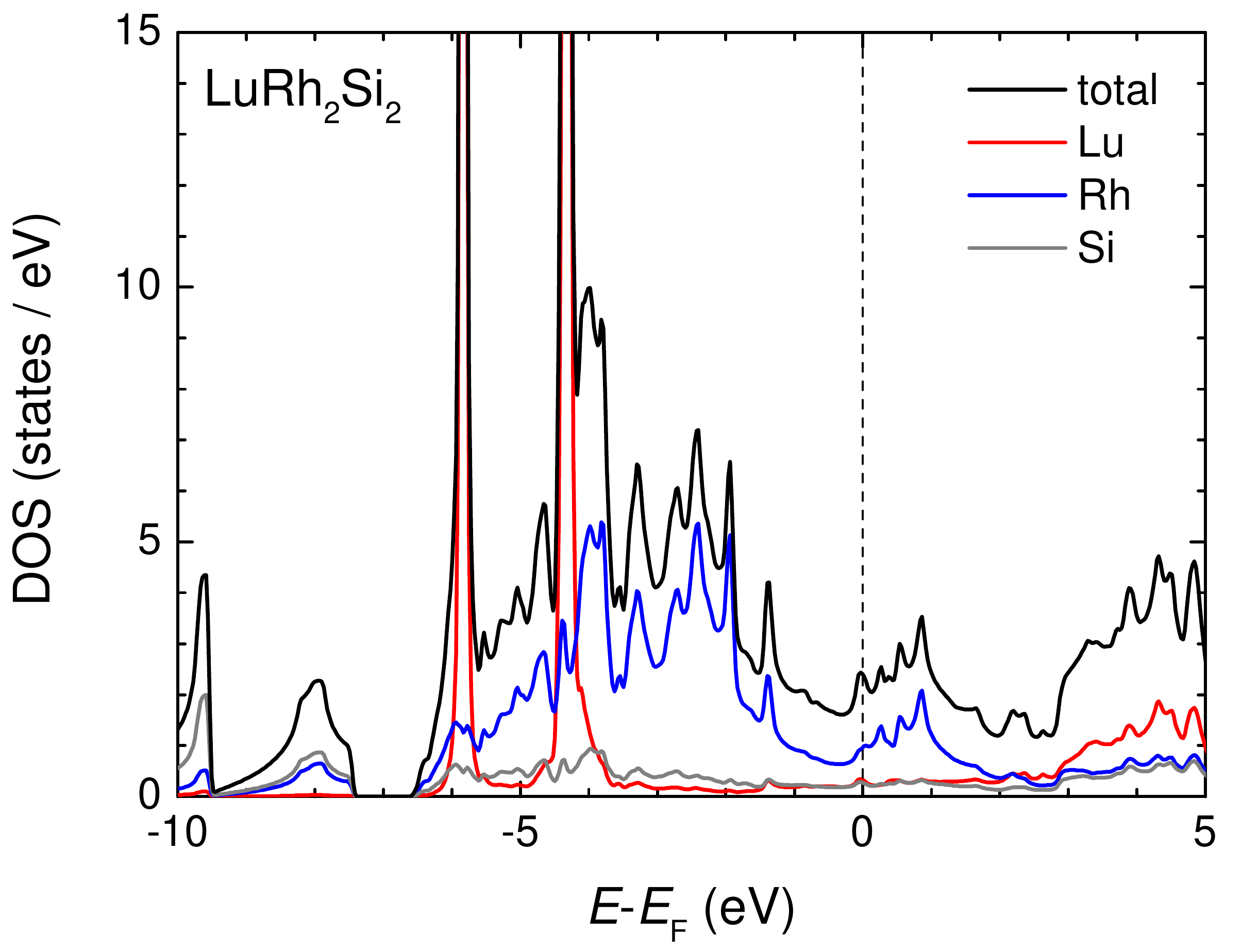}%
\caption{Density of States plotted against energy with the Fermi energy \EF\ as reference. Colors represent the band character; at the Fermi energy Rh 4$d$ states dominate. }%
\label{fig:DOS}%
\end{figure}

We now turn to the DOS as displayed in Fig.~\ref{fig:DOS}. Most prominently, the large DOS peaks well below the Fermi energy at \SIlist{-6;-4}{\electronvolt} arise from the Lu 4$f$ states, in good agreement with expected behaviour for a completely filled $f$ shell with small radial extension. 
The Rh $d$ states originating from the large overlap of Rh orbitals within the Rh-Si layers (cf.\ crystal structure in inset of Fig.~\ref{fig:totEn}) appear to be distributed over a large energy range and dominate the DOS at the Fermi energy. A minor contribution arises from the Si states as part of the Rh-Si layers. The admixture of the Si states within this layer causes the sensitivity to the Si position, as will be discussed below \cite{Reiss2013}. The Si $s$ states lie far below the Fermi energy at about \SI{-10}{\electronvolt}. The total DOS at the Fermi energy amounts to 
\SI{2.4}{states\per(\electronvolt.unit cell)} which is slightly higher than reported previously (based on a different \zSi) \cite{Friedemann2010c}, corresponding to a Sommerfeld coefficient $\gamma \approx \SI{5.7}{\milli\joule\per\kelvin\squared\per\mole}$ in  good agreement with the value of $\gamma \approx \SI{6.5}{\milli\joule\per\kelvin\squared\per\mole}$ found experimentally \cite{Friedemann2010c}.

\begin{figure}%
\includegraphics[width=\columnwidth]{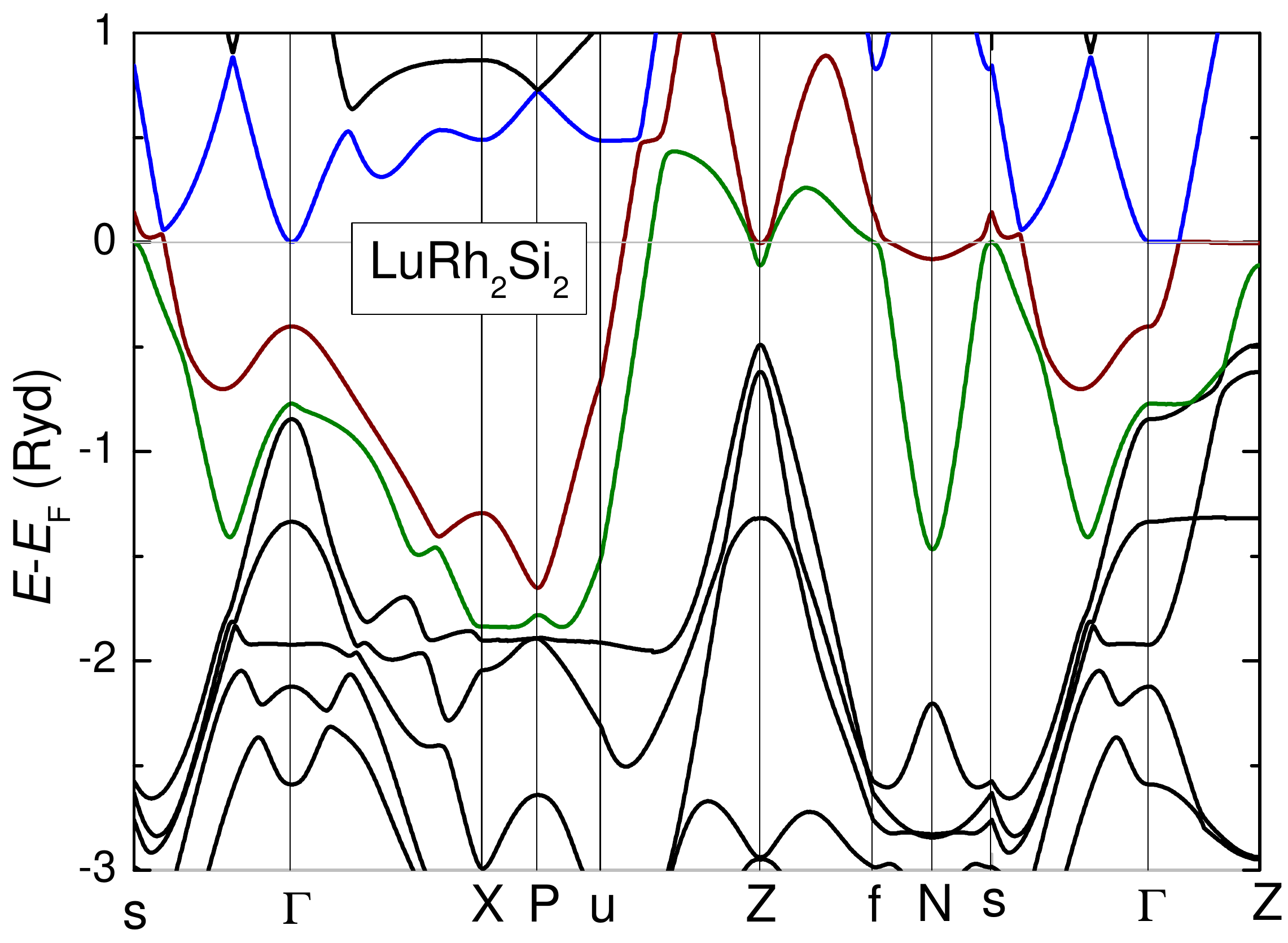}
\caption{Band Structure of \LRS\ for $\zSi = \SI{0.379}{c}$ along symmetry lines with the Fermi energy $E_{\text F}=0$ as the reference energy. The conventional notation is adopted with $Z (0,0,1)$, $\Gamma (0,0,0)$, $X (1,1,0)$, $P (1,1,1)$ and $N (1,0,1/2)$ in units of $(\pi/2a,\pi/2a,\pi/2c)$. The labels $s$,$f$, and $u$ refer to $(\tilde{a},0,0)$, $(\tilde{b}, 0, 1)$, and $(\tilde{b},\tilde{b},1)$, respectively, with $\tilde{a} = 1+(a/2c)^2$ and $\tilde{b}= 1-(a/2c)^2$. Bands close to the Fermi energy are represented in colour. }%
\label{fig:BS}%
\end{figure}

The electronic band structure is shown in Fig.~\ref{fig:BS}. In general, we find good agreement with previously reported band structure calculations on \LRS\ and \YRS. Small differences in the vicinity of the Fermi energy of our band structure calculations with respect to previous studies are due to the strong sensitivity to the Si position%
\cite{Rourke2008,Friedemann2010c,Wigger2007,Jeong2006}.

As discussed above, most of the bands have dominantly Rh 4$d$ character. Around the $\Gamma$ point 
the Si 3$p_z$ states admix strongly, and this causes the sensitivity of the band structure to \zSi\ \cite{Reiss2013}. Changes in the Si position change the penetration of  the Si $p_z$ orbitals into the Rh~$d_{x^2-y^2}$ orbitals within the Rh layers. This causes a shift of the bands at the Fermi energy around the Z point as detailed in Ref.~\onlinecite{Reiss2013}. As a consequence, the character of the band intersecting the Fermi energy changes from Rh~$d_{x^2-y^2}$ like at low \zSi\ to Si $p_z$ like at $\zSi \geq \SI{0.379}{c}$.  The mass of the bands associated with the Rh $d_{x^2-y^2}$ and the Si $p_z$ states differ significantly, as can be seen from the stronger curvature of the upper band. This will serve as an important point in comparison with experimental data below.


\begin{figure}%
\includegraphics[width=.47\columnwidth]{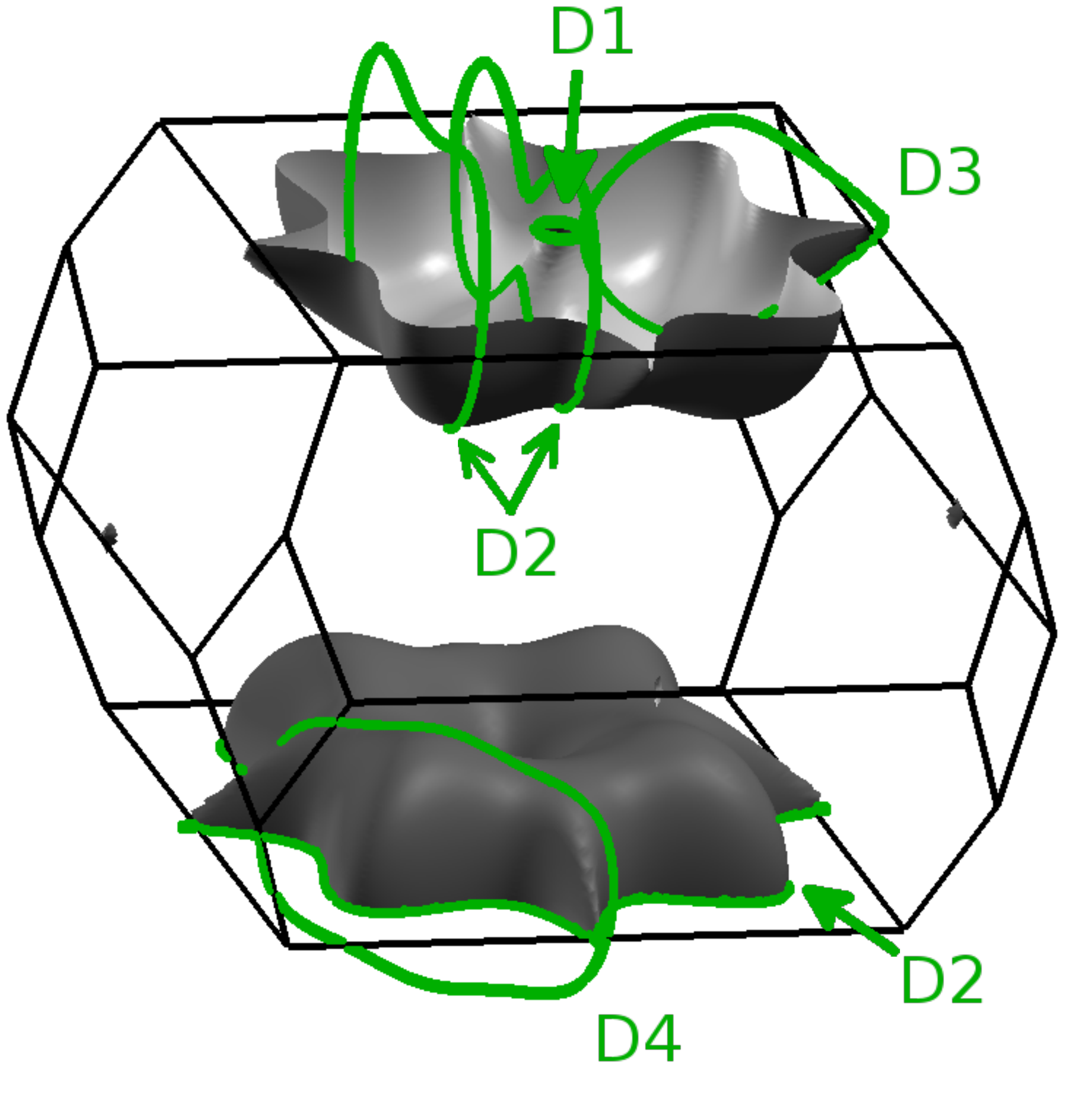}%
\includegraphics[width=.53\columnwidth]{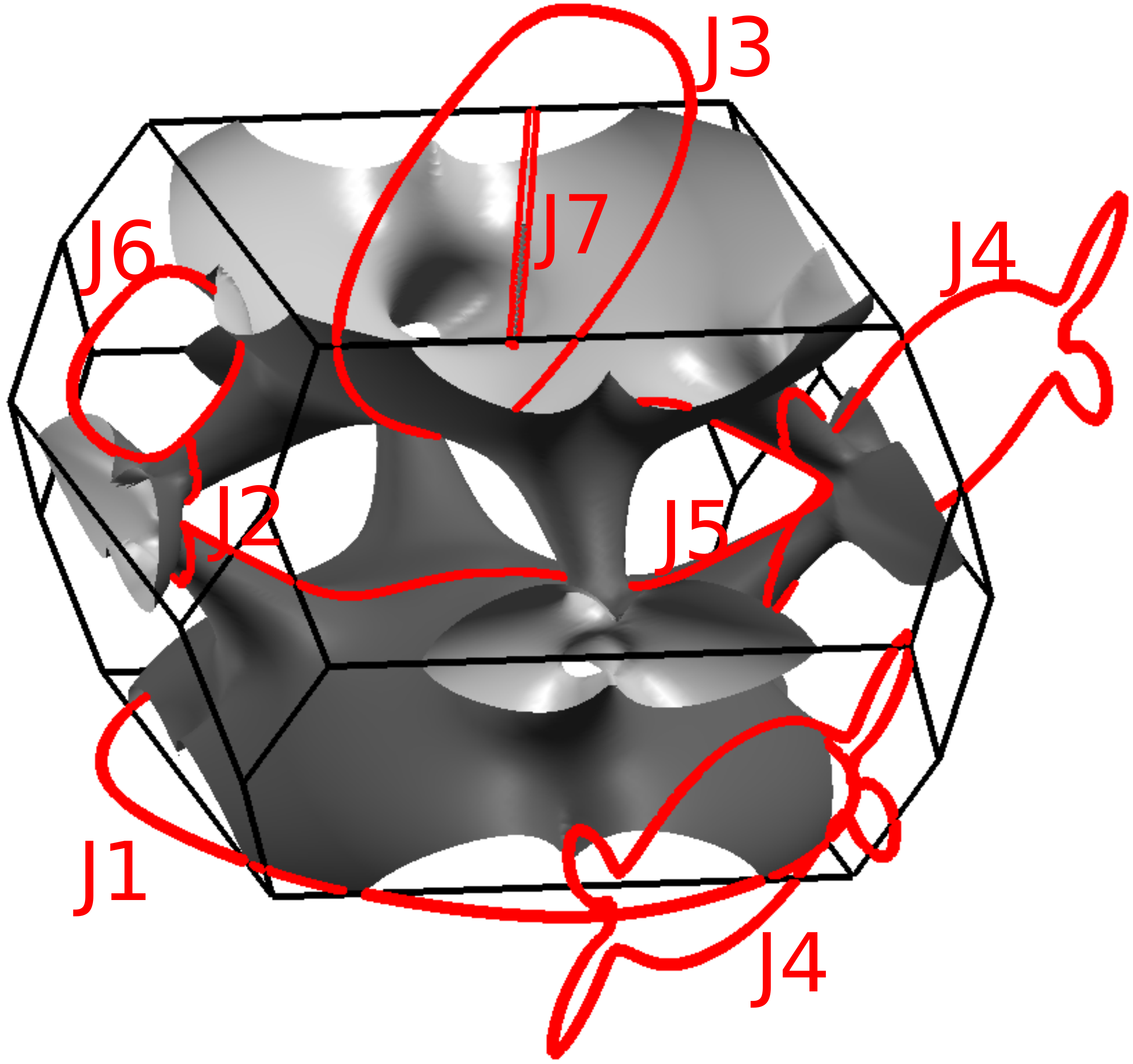}%
\caption{Calculated Fermi surfaces: A doughnut shaped D sheet and a interconnected ''jungle gym`` J sheet were obtained using the experimental value for the Si position $\zSi=\SI{0.379}{c}$. Green and red lines represent selected extremal orbits.}%
\label{fig:FS}%
\end{figure}

Three bands are close to the Fermi energy (cf.\ coloured bands in Fig.~\ref{fig:BS}). For $\zSi=\SI{0.379}{c}$, however, only the two at lower energy (green and red) cross the Fermi energy. These two bands give rise to the Fermi surface sheets shown in Fig.~\ref{fig:FS}, nicely resembling 
previous calculations \cite{Wigger2007,Rourke2008,Rourke2009,Friedemann2010c}. They consist of two $Z$-centred surfaces, a closed doughnut shaped D sheet and a J sheet (previously dubbed ''jungle gym``), with the latter one extending across the Brillouin zone boundary, connected via tubes along the $a^{\star}$ direction.

The main differences to previous band structure calculations arise due to the corrected Si position of our band structure calculations \cite{Reiss2013}: We do not obtain a third pillbox shaped P sheet. This can be seen from the band structure plot  in which the band associated with the pillbox (blue in Fig.~\ref{fig:BS}) remains slightly above the Fermi energy at the $\Gamma$ point. This P band is only populated for $\zSi \leq \SI{0.378}{c}$. 
At $\zSi = \SI{0.379}{c}$, we find a very thin central pillar in the J sheet (encircled by orbit J7 Fig.~\ref{fig:FS}) which arises from a hybridization with the states forming the P band. The fact that the pillar is present but the P sheet is absent is due to a small gradient in the dispersion relation from the $\Gamma$ point to the Z point as detailed in Ref.~\onlinecite{Reiss2013}. In fact, the pillar is disconnected from the main sheet due to this gradient, as can be seen in Fig.~\ref{fig:FS}.


 The hole in the D sheet persists even for large variations of the Si position \zSi\ as shown in Ref.~\onlinecite{Reiss2013}. This hole is associated with the band crossing the Fermi energy around the Z point. It is minimal in size for $\zSi = \SI{0.379}{c}$ and widens for both decreasing and increasing \zSi. The character of the hole in the D sheet, however, changes from Rh $d_{x^2-y^2}$ to Si $p_z$ for $\zSi \gtrsim \SI{0.379}{c}$. This is accompanied by a change in the mass of this orbit as mentioned above. The steep curvature of the Si $p_z$-like band corresponds to a lower mass than the flat Rh $d_{x^2-y^2}$ band. We shall use these sensitivities of the Fermi surface topology and the masses of extremal orbits for a detailed comparison with the observed quantum oscillations below.

%
%
\section{\SdH\ Measurements}
\subsection{Experimental Details}
Single crystal samples of \LRS\ were grown in indium flux as described earlier \cite{Maquilon2007}.
Shubnikov-de Haas oscillations on a bar shaped sample were measured using a standard four probe resistivity measurement. Contacts were provided by \SI{25}{\micro\meter} gold wires spot welded to the sample of dimensions of approximately $20 \times 100 \times \SI{2000}{\cubic\micro\meter}$. The current was applied within the basal plane at an angle of $\approx \SI{10}{\degree}$ from the (100) axis. Measurements were performed in a $^3$He/$^4$He-dilution refrigerator in magnetic fields $B$ up to \SI{16}{\tesla}. The oscillatory part was deduced by subtracting a polynomial fit from the raw data. The order of polynomial background was choosen such that its subtraction does not interfere with the lowest frequencies. Oscillation frequencies were determined after Fourier transformation. In order to deduce the angular dependence of the oscillation frequencies the magnetic field was rotated within the crystallographic basal plane and from the (100) direction towards the (001) direction. 

Quantum oscillation frequencies are related to the extremal cross-sectional area $A$ of the Fermi surface via the Onsager relation $F=\hbar A/2\pi e$. For simple non-magnetic metals the damping of quantum oscillations is captured by the Lifshitz-Kosevich formula with the damping factors due to impurity scattering \cite{Schoenberg2009}
\begin{equation}
	R_{\text D} = \exp\left(-\frac{B_{\text D}}{B}\right) \quad \text{with} \quad 
	B_{\text D} = \sqrt{\frac{e \hbar^3 F}{2\pi} }\frac{\kappa p}{k_{\text B}l_0 }
\label{eq:Dingle}
\end{equation}
and thermal broadening of the Fermi-Dirac distribution
\begin{equation}
 R_T = \frac{X}{\sinh{X}} \quad \text{with} \quad X= \kappa p \frac{T }{B} \frac{m^{\star}}{m_{\text e}}
\label{eq:mass}
\end{equation}
Here, $\kappa = 2\pi^2 k_{\text B}m_{\text e} / e \hbar \approx \SI{14.7}{\tesla \per\kelvin}$ while $p$ denotes the index of the harmonic, i.e. $p=1$ for fundamental frequencies.
These two damping factors can be seen in the field and temperature dependence, respectively, of the amplitude of a single frequency. By fitting eqs.~\ref{eq:Dingle} and \ref{eq:mass} we extract the mean free path $l_0$ and the effective mass $m^{\star}$ of the charge carriers. The large temperature and field range covered in our experiments yield high accuracy determinations of these parameters for the most prominent oscillation frequencies. 
%
%
\subsection{Experimental Results}
\begin{figure}%
\includegraphics[width=\columnwidth]{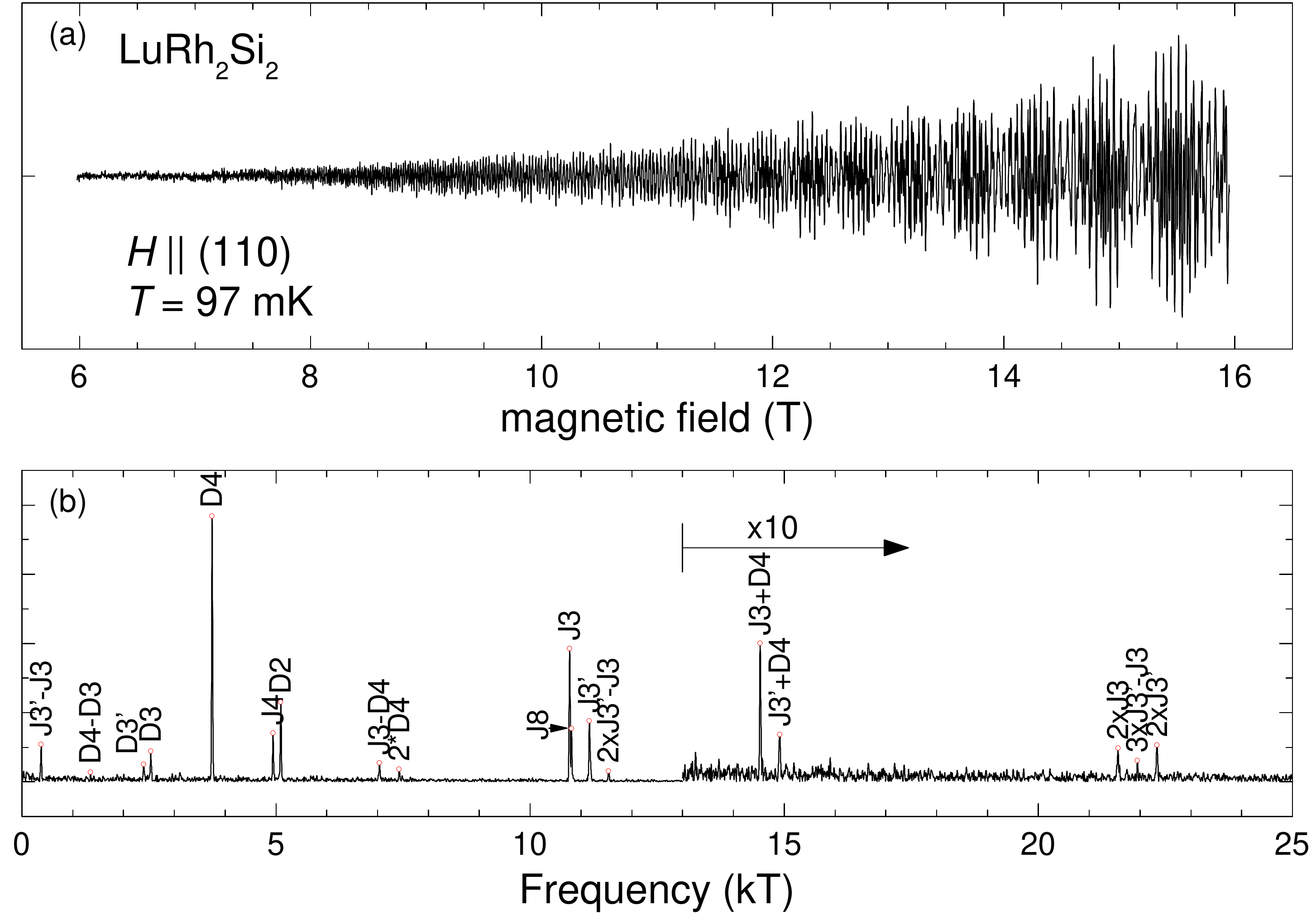}%
\caption{Shubnikov-de Haas oscillations. (a) Oscillatory part of the resistivity between \SI{6}{\tesla} and \SI{16}{\tesla} for field along the (110) direction measured at \SI{100}{\milli\kelvin}. (b) Fourier transform power spectrum showing quantum oscillation frequencies. Data above \SI{13}{\kT} are magnified by a factor of 10 for better visibility of the high frequency peaks.}%
\label{fig:SdH_Dingle}%
\end{figure}

Figure \ref{fig:SdH_Dingle} shows a representative trace of the oscillatory part of the resistivity taken at \SI{100}{\mK} for fields between \SIlist{6;16}{\tesla} with the field applied along the (110) direction. From the Fourier transformed power spectrum, numerous oscillation frequencies are resolved with signal to noise ratios exceeding 100. Note that data in the Fourier spectrum above \Freq{13} are multiplied by a factor of 10 in order to make the high-frequency peaks more visible. In total, 19 frequencies are detected for this field orientation, some of which are identified as harmonics and some arise from mixing of fundamental frequencies as discussed below. The nomenclature of the frequencies reflects the assignment to orbits on the different Fermi surface sheets, which we deduce from the comparison with band structure calculations  below.

\begin{figure}%
\includegraphics[width=.9\columnwidth]{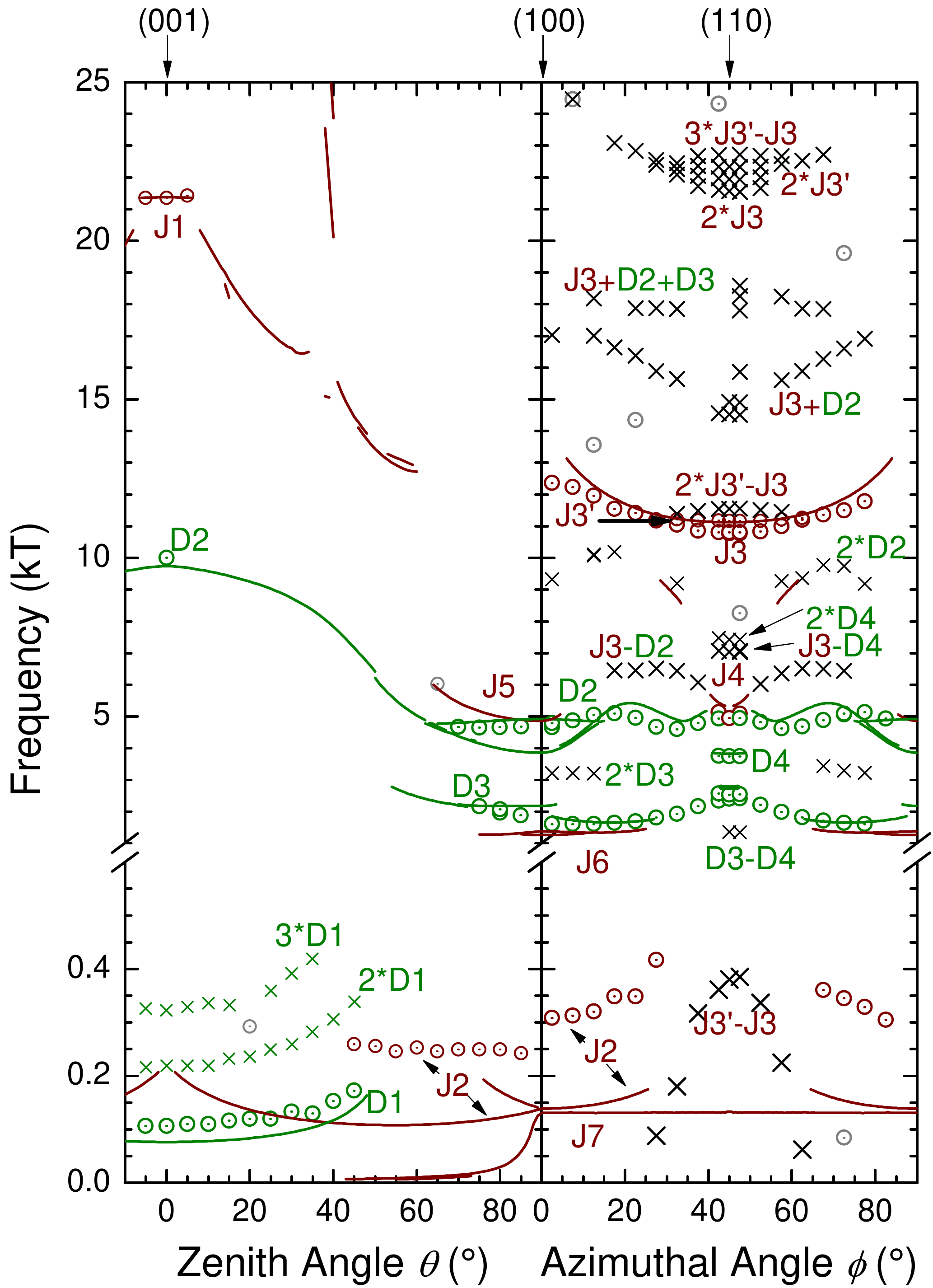}%
\caption{Angular dependence of oscillation frequencies. Data below \SI{0.45}{\kT} are displayed on an expanded scale. No oscillations are observed in the omitted range between \SI{0.45}{\kT} and \SI{1}{\kT}. Open circles mark fundamental frequencies, whereas crosses correspond to harmonics and frequencies which were identified to arise from mixing of fundamental frequencies (see text for details, Sec.~\ref{subsubsec:Mixing}). Solid lines represent calculated frequencies from our band structure calculations with green and red representing the D and J sheet, respectively, labels refer to the orbits depicted in Fig.~\ref{fig:FS}.
Top labels and arrows indicate crystallographic orientations.}%
\label{fig:AngDep}%
\end{figure}

The angular dependence of the oscillation frequencies is shown in Fig.~\ref{fig:AngDep} for fields rotated from the (001) direction towards the (100) direction (left panel) and further within the basal plane (right panel). The symmetry observed around \SI{45}{\degree} in the right panel nicely reflects the crystallographic symmetry around the (110) direction. We observe frequencies from \SIrange{0.06}{42.7}{\kT} (for fields along (001))%
. Generally, more frequencies are observed for fields within the basal plane. 

By comparison with our band structure calculation we identify the fundamental frequencies and assign them to extremal orbits of the Fermi surfaces as shown in Fig.~\ref{fig:FS}. The angular dependence of the predicted frequencies deduced from calculations utilizing $\zSi = \SI{0.379}{c}$ are included in Fig.~\ref{fig:AngDep} as solid lines. 

\begin{table*}
	\centering
	\caption{Frequencies and cyclotron masses for field orientations along (110),(001), and (100).
		\label{tab:Masses}}
		\begin{tabular}{r@{\hspace{1em}}ll cc @{\hspace{1em}} ll c @{\hspace{2em}} r@{\hspace{1em}}ll c @{\hspace{1em}} ll }
			\hline\hline
			\multicolumn{7}{l} {(110)} && \multicolumn{6}{l} {(001)} \\ 
			\cline{1-7} \cline{9-14} 
			& \multicolumn{3}{c}{experimental} && \multicolumn{2}{c}{calculated} &&& \multicolumn{2}{c}{experimental} && \multicolumn{2}{c}{calculated}\\
			\cline{2-4} \cline{6-7} \cline{10-11} \cline{13-14}
			orbit & $F (\si{\kilo\tesla})$ & $\mstar / m_{\text e}$ & $l_0$ (\si{\micro \meter}) && $F (\si{\kilo\tesla})$ & $\mstar / m_{\text e}$ & &
			orbit & $F (\si{\kilo\tesla})$ & $\mstar / m_{\text e}$ && $F (\si{\kilo\tesla})$ & $\mstar / m_{\text e}$ \\ 
			
			\cline{1-4} \cline{6-7} \cline{9-11} \cline{13-14} 
			J7 &&&&&
			0.13 & 4.3 
			 &&
			D1 & 0.11 &	0.11(5) && 
			0.08 & 0.11	\\ 
			

			$\Jp - \text J3$ & 0.38	& 2.5(4)& 0.15(2) & &
			 & & & 
			$2 * \text{D1}$ & 0.22	& 0.2(1) \\
			{D3'} & 2.41 &	0.7(3) &&&
			 2.75 & 0.4  && 
			D2 & 9.99	& 2.4(3) &&
			9.7 & 1.9 \\
			D3 & 2.54 &	1.3(5) &&&
			&&&
			J1 & 	21.35	& 1.54(2) &&
			21.4 & 1.3 \\
			
			D4 & 3.75 &	0.82(2) &0.32(2) &&
			3.8 & 0.7 &&
			$2*\text{J1}$ & 42.73 &	2.5(5) \\
			
			J4 & 4.95 &	1.4(5) &&&
			5.3 & 4(2) \footnote[4]{small angular deviation induces large changes in the mass}&&
\\
			
			D2 & 5.10	 &  1.17(3) &0.40(5)&&
			5.2 & 1 \footnote[5]{extremal orbit only predicted at \SI{5}{\degree} deviation from (110) direction} &&
			(100)\\
			\cline{9-14}
			$\text{J3}-\text{D4}$ & 7.05 & 	 2.3(1)& 0.25(5) &&&&&
			& \multicolumn{2}{c}{experimental}&& \multicolumn{2}{c}{calculated}\\
			
			\cline{10-11} \cline{13-14}
			$2*\text{D4}$ & 7.43 &	1.6(3) &&&&&&
			orbit & $F (\si{\kilo\tesla})$ & $\mstar / m_{\text e}$ && $F (\si{\kilo\tesla})$ & $\mstar / m_{\text e}$\\
			\cline{9-11} \cline{13-14}%
			
			J3 & 10.78  &  1.3(2) \footnote{The mass of \SI{1.3}{\me} was deduced from the analysis of mixed frequencies (see text), The measured mass of \SI{2.29(3)}{\me} is compromised by the J8 frequency which could not be separated in the mass study.}
			&0.4(1)&&
			11.1 & 1.1 &&
						 J7 &&&&
			 0.13 & 4.3 \\
			J8 & 10.81 &&&&&&&
			J2 & 0.25 &	1.3(5) \footnotemark[1]{} &&
			0.14 & 0.5 \footnotemark[1]{} \\
						\Jp & 11.17 &	1.26(2) &0.3-0.6 \footnote{field dependent mean free path, cf.\ Fig.~\ref{fig:Dingle} and text} &&
			&&&
			J6 &&&&
			 1.26 & 1.8 \\			

			$2*\Jp - \text J3$ & 11.54 &	3.1(3) &0.38(4)&&&&&
			J6' &&&&
			 1.38 & 2.6 \\
			$\text{J3} + \text{D4}$ & 14.53 &	2.21(7) &0.33(3)&&&&&
						D3 & 1.61 &	0.5(1) \footnote[2]{$\approx \SI{3}{\degree}$ off towards (110)}&&
			1.8 & 0.56 \footnotemark[2]{}\\
			$\Jp + \text{D4}$  & 14.91 &	1.7(2) &&&&&&
						D3 & 1.81 & 0.4(3) \footnotemark[1]{} &&
			2.2 & 0.5 \footnotemark[1]{} \\
			$2* \text{J3}$ & 21.57 &	3.1(3) &&&&&&
			$2*\text{D3}$ & 3.22	& 1.2(4) \\
			$\Jp + \text J3$ & 21.95	& 	2.4(5) &&&&&&
			D2 & 4.64	& 0.8(1) \footnote[1]{$\approx \SI{5}{\degree}$ off towards (001)}&&
			4.9 & 0.6 \footnotemark[1]{}\\
			$2 * \Jp $ & 22.33 &	2.6(2) &&&&&&
			D2 & 4.64 &	1.1(1) \footnotemark[2]{}
			 &&
			 4.9 & 0.6 \footnotemark[2]{}\\
			 &&&&&&&& 
			D2 & 4.74 &	1.1(1) \footnotemark[2]{} \\
			 &&&&&&&& 
			 J3 & 12.36 & 2.43(2)\footnotemark[2]{}  
			 && 12.7 & 3 \footnotemark[2]{}\footnotemark[5]{}\\
			\hline\hline
		\end{tabular}
\end{table*}

For the rotation of the magnetic field from the (001) to the (100) direction the assignment is unambiguous. First, an orbit D1 associated with the hole in the D sheet is predicted which nicely matches the angular dependence of an observed branch. In particular, the angular range over which this D1 is observed nicely agrees with the predicted angular dependence. The difference between the observed and predicted frequency is very small -- note the amplified low frequency scale in the lower part of Fig.~\ref{fig:AngDep}
. The experimentally determined mass of this orbit agrees very well with the predicted mass as can be seen from Tab.~\ref{tab:Masses}. In fact, the mass of this orbit strongly supports the usage of $\zSi =\SI{0.379}{c}$, where our band structure calculations predict a Si $p_z$ character of the states forming this orbit. For smaller \zSi\ the dominating Rh $d_{x^2-y^2}$ band character and flatter dispersion relation yield a significantly larger mass \cite{Reiss2013}. We also detect the second and third harmonic of the D1 orbit with the mass of the second harmonic very precisely being double of that of the fundamental frequency (the mass of the third harmonic could not be determined due to the strong reduction in intensity).

It is unlikely that the observed frequency corresponds to the pillbox shaped Fermi surface sheet predicted at $\zSi \leq 0.378$. An outer orbit along the convex shape of the pillbox would lead to a $1/\cos\theta$  corrected by a small reduction as $\theta \gtrsim \SI{45}{\degree}$ associated with a rounding of the corners of the pillbox. The experimental data, however, show a small excess with respect to a $1/\cos\theta$ form which is in good agreement with the inner orbit  in the convex shaped D sheet. Furthermore, the P sheet has extremal orbits extending all the way to $\theta = \SI{90}{\degree}$ and also predicts a constant frequency for rotations in the basal plane which is not observed. In general, it is risky to draw conclusions from failure to observe quantum oscillations, particularly since the amplitude of the pillbox orbits is expected to be reduced as the angle differs significantly from axial direction. However, the fact that we observe a strong signal including the 2nd and 3rd harmonics up to $\theta = \SI{50}{\degree}$ which suddenly disappears at larger angles is in contrast to the continuous reduction expected for a pillbox. By contrast, this agrees well with the expectations for the inner orbit in the D sheet, which is not present for angles larger than $\theta \leq \SI{50}{\degree}$.

Frequencies detected at \SI{9.99}{\kT} and \SI{21.35}{\kT} for field along (001) can be assigned to the orbit D2 -- the circumference of the D sheet -- and the orbit J1 -- the circumference of the J sheet. In both cases we find a good agreement of predicted and experimentally determined effective masses as can be seen from Tab.~\ref{tab:Masses}. In addition, we detect the 2nd harmonic of the J1 orbit with its mass in agreement with twice the mass of the fundamental frequency. For the J1 orbit we also see a good agreement with the predicted angular dependence matching very nicely the range over which this orbit is predicted to be extremal. In fact. the predicted range is very small due to the arms of the J sheet interrupting this orbit for larger angles. Consequently, our quantum oscillation results strongly suggest that these arms are present in the J sheet. Like for the D1 orbit it is unlikely that the J1 orbit is lost due to a to small signal at angles $\theta >\SI{5}{\degree}$, as we observe a very strong signal including the 2nd harmonic, which suddenly vanishes. The D2 frequency, by contrast, has a very small amplitude for fields parallel (001) and may very well be lost due to further reduction as the field is rotated away from this direction.

As the field direction approaches the basal plane, the D2 orbit is detected again at a frequency of \SI{4.7}{\kT} for $\theta \leq \SI{70}{\degree}$ nicely reproducing the flat angular dependence of one branch associated with this orbit as well as the mass expected for the (100) direction (cf.\ Tab.~\ref{tab:Masses}). The two branches correspond to different extremal orbits, as depicted in Fig.~\ref{fig:FS}. This orbit is also observed for rotation in the basal plane with a good match to the predicted angular dependence. Only in the very vicinity of the (110) direction of magnetic field is  this orbit not expected to persist. However, for angles $\SI{42.5}{\degree}\leq\phi \leq\SI{45}{\degree}$ two frequencies \Freq{4.9} and \Freq{5.1} are observed. One of these may be associated with the J4 orbit of the J sheet while the other one might still arise for the D2 orbit due to small misalignment. In fact, the mass predicted for the D2 orbit at a small angle nicely matches the observed mass of the \Freq{5.1} oscillations (cf.\ Tab.~\ref{tab:Masses}). The mass of the J4 orbit is expected to be much larger then observed  for the \Freq{4.9} frequency, however again, small misalignments yield an improved agreement.

A frequency of about $\SI{1.6}{\kT}$ is observed close to the (100) direction which is assigned to an orbit D3 circling from the hole to the circumference of the D sheet (cf.\ Fig.~\ref{fig:FS}). Both the angular dependence and the mass agrees well with the results of the band structure calculations (cf.\ Tab.~\ref{tab:Masses}) for field orientations out of the basal plane and along the (100) direction. For intermediate angles in the basal plane $\SI{30}{\degree} \leq \phi \leq \SI{40}{\degree}$ this orbit is expected to be non-extremal. It might be possible that we observe oscillations associated with a non-extremal orbit. This is in agreement with the fact that the amplitude of this frequency is strongly reduced to almost noise level in this angular range while other frequencies preserve a strong amplitude. However, already small misalignments in the experiment or approximations in the band structure calculations can result in an extremal orbit. It would require further experimental and computational work to scrutinize the hypothesis of a non-extremal orbit.

For angles close to the (110) direction $\SI{40}{\degree} \leq \phi \leq \SI{45}{\degree}$, two frequencies D3 and D$3^{\prime}$ are observed. Both have almost no angular dependence in agreement with an extremal orbit D3 present over a very limited angular range $\SI{43}{\degree} \leq \phi \leq \SI{45}{\degree}$. The mass of this this orbit agrees within experimental accuracy with the lower frequency D3. For reduced \zSi\ we also find two orbits D3 and D$3^{\prime}$ with frequencies close by. Their masses resemble those of the two detected frequencies. This supports the above suggestion of corrections to the band structure calculations to improve agreement with experimental results.

The frequency with the largest amplitude in the power spectrum at \Freq{3.7} (cf.\ Fig.~\ref{fig:SdH_Dingle}) is observed in a small angular range of \SI{2.5}{\degree} around the (110) direction only. This nicely matches the predicted range for the D4 orbit of the D sheet with also the predicted mass in good agreement with the observed value (cf.\ Tab.~\ref{tab:Masses}). This orbit is very sensitive to changes in the Si position \zSi\ in our band structure calculations and vanishes for $\zSi > \SI{0.380}{c}$ giving a strong upper boundary for \zSi.

A frequency with small angular dependence around $\SI{0.25}{\kT}$ is observed in the rotation study from (001) to (100) in the range $\SI{45}{\degree} \leq \theta \leq \SI{90}{\degree}$. This frequency is slightly larger than that expected for the J2 orbit circling on the outside of the arms in the J sheet but roughly matches its angular dependence. In addition, this orbit is predicted to be continued for rotations in the basal plane up to $\phi \leq \SI{25}{\degree}$ with a significant increase of frequency as $\phi$ increases. This nicely matches the continued branch observed (the small offset of \SI{0.05}{\kT} from rotation out of the basal plane to that in the basal plane may be due to a small misalignment of the sample in the two subsequent rotation studies). 

A group of strong frequencies is found around \Freq{11} for fields along the (110) direction. The lowest  of these frequencies (\Freq{10.8}) has the largest amplitude in the power spectrum (cf.\ Fig.~\ref{fig:SdH_Dingle}) and can be traced over the complete range of our rotation study towards the (100) direction down to $\phi \geq \SI{2.5}{\degree}$. A high resolution study of this frequency over a wide field range reveals a two peak structure (cf.\ Fig.~\ref{fig:SdH_Dingle}) with \Freq{10.78} and \Freq{10.81}. The higher amplitude arises from the lower frequency, i.e. \Freq{10.78}. The angular dependence of this frequency shows good agreement with the predicted angular dependence of the J3 orbit circling the inside of the main body of the J sheet. It is natural to assign the frequency at \Freq{10.78} with larger amplitude to the J3 orbit. We surmise that the the frequency of \Freq{10.81} with the lower amplitude is a secondary effect and label this frequency J8 although we have no proof for it to be related to the J sheet.

The fact that J3 is detected almost all the way to the (100) direction indicates that the arms of the J sheet are very small as these interrupt the J3 orbit for fields along (100). In fact, for smaller \zSi\ the band structure calculations predict the angular range of the J3 orbit to narrow very rapidly, thus, the observed angular dependence of the J3 orbit strongly supports the experimental value for \zSi.
 The frequencies at \Freq{11.2} and \Freq{11.5} have a smaller amplitude and are limited in the angular range to $\SI{27.5}{\degree}\leq \phi \leq \SI{45}{\degree}$. 
The frequencies around \Freq{11} give rise to various harmonics and mixed frequencies. From the detailed analysis we identify  one of the \Freq{10.8} frequencies and the \Freq{11.2} frequency to be the fundamental frequencies, we dub them J3 and \Jp\ as both bear resemblance with the angular dependence and masses expected for the J3 orbit.

\begin{figure}%
\includegraphics[width=.5\columnwidth]{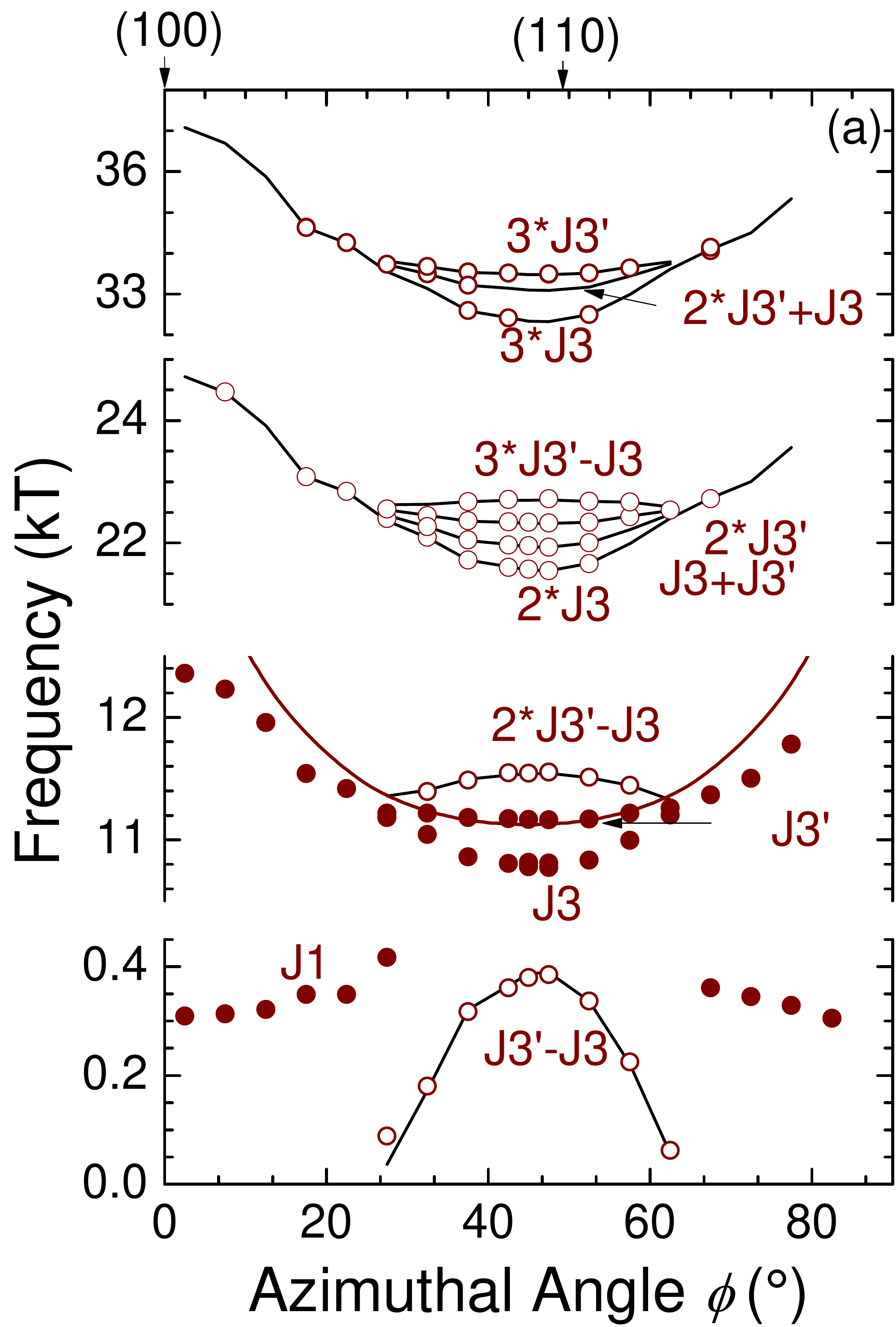}%
\hfill
\includegraphics[width=.5\columnwidth]{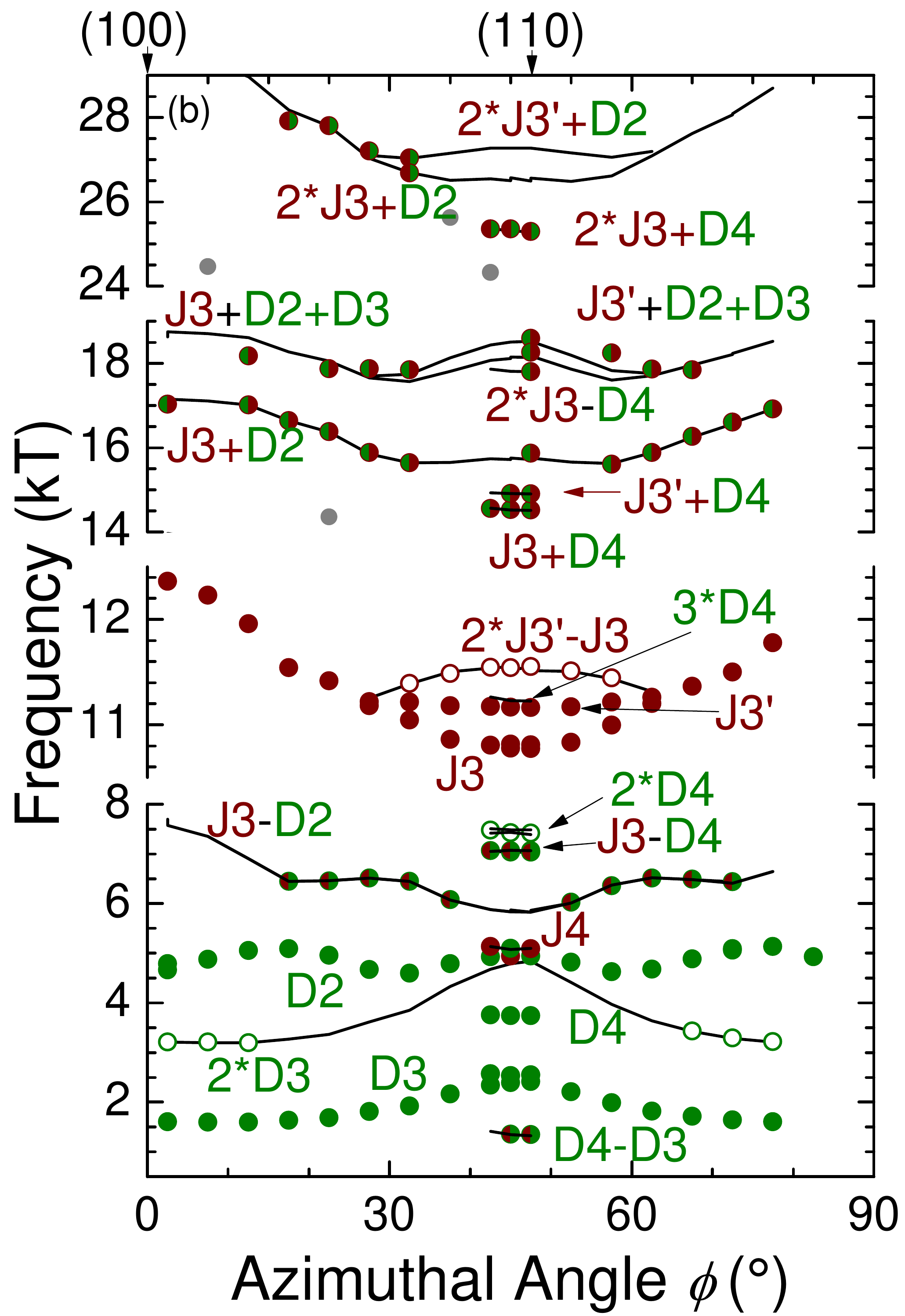}%
\caption{Frequency mixing was identified with the help of an algorithm searching for sums and differences of frequencies. Fundamental frequencies were deduced from the comparison with band structure calculations. Symbols mark detected frequencies, solid black lines mark frequencies calculated from sums and differences of fundamental frequencies as indicated by the labels. (a) depicts the mixing of the J3 orbit with the \Jp\ orbit and harmonics while (b) depicts the mixing and harmonics of the J3 and \Jp\ orbit with orbits of the D-sheet, i.e., D2, D3, and D4. Solid red lines in (a) reproduce the calculated angular dependence of the J3 orbit. 
}
\label{fig:Mixing}%
\end{figure}

\subsubsection{Frequency Mixing Analysis}
\label{subsubsec:Mixing}
The measured frequencies are presented in Fig.~\ref{fig:Mixing} (a) together with the calculated harmonics and mixed frequencies of J3 and \Jp.
Clearly, we detect second and third harmonics of both J3 and \Jp. We also detect the sum and difference of these two frequencies, i.e. $\text J3 \pm \Jp$. In addition, we detect the sum and difference of twice \Jp\ with J3, i.e. $2*\Jp\pm \text J3$. Finally, we also detect the difference of $3*\Jp-\text J3$. The analogous sum $3*\Jp+\text J3$ was not detected, possibly because of unfavourable sampling for such high frequencies.

Possible mechanisms for frequency mixing are magnetic breakdown and magnetic interaction. The former arises when the magnetic field exceeds the equivalent of the energy gap associated with orbits close by in $k$-space \cite{Schoenberg2009}. The latter arises from an oscillating magnetisation that is comparable to the external magnetic field and has for instance been observed in gold \cite{Schoenberg2009}.

The pattern of frequencies, i.e., the observation of sums and differences is most consistent with 
magnetic interaction of two fundamental frequencies J3 and \Jp. In particular, considering first and second order effects, i.e. including fundamental frequencies J3 and \Jp\ and second harmonics $2*\text J3$ and $2*\Jp$ in the oscillating magnetisation yields the observed second and third harmonics as well as the sums and differences of the fundamental and harmonic frequencies. 

\begin{figure}%
\includegraphics[width=.8\columnwidth]{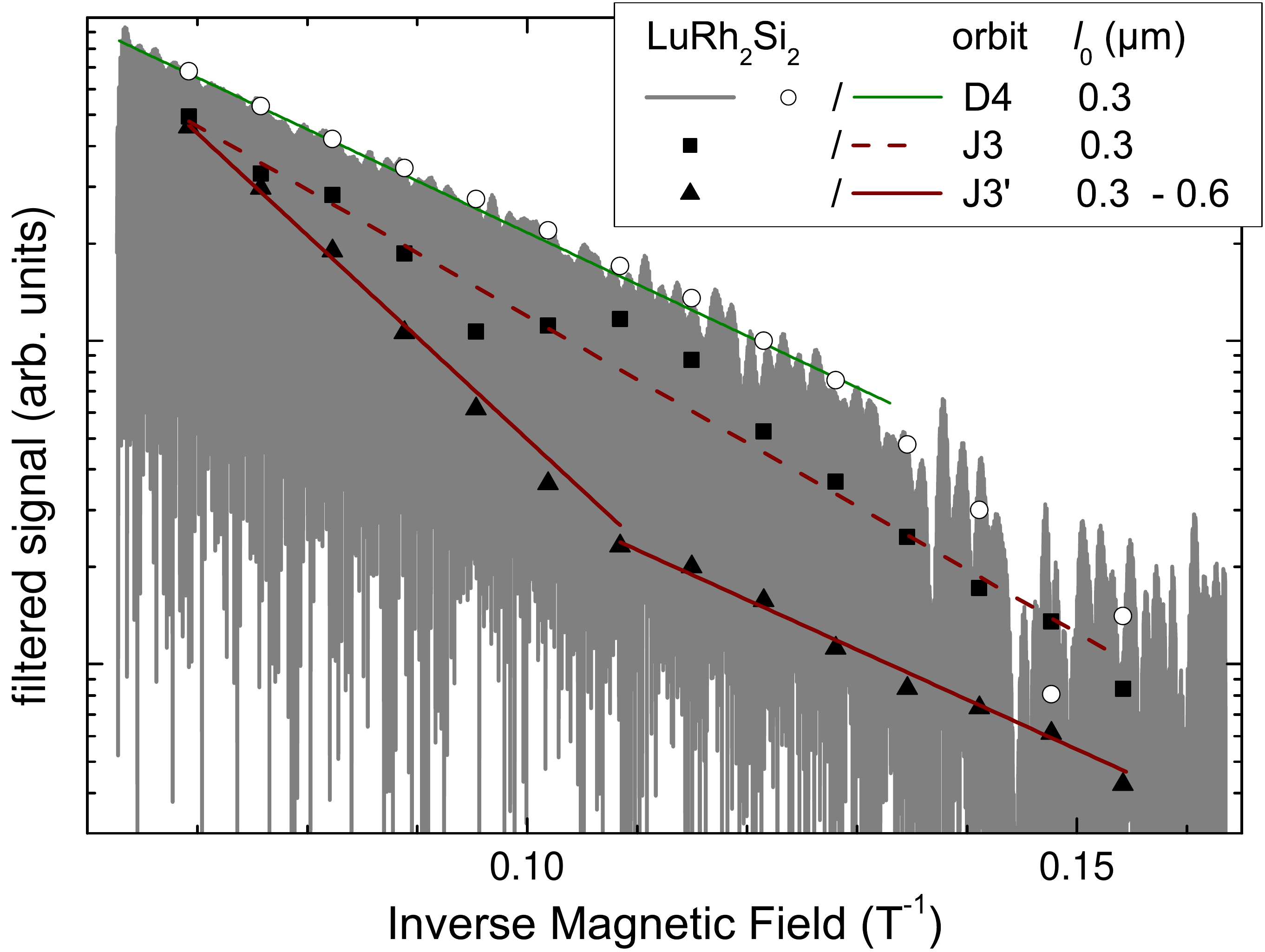}%
\caption{Dingle analysis of the field dependent amplitude \cite{Schoenberg2009} for the frequencies D4, J3 and \Jp. The mean free path was extracted by fitting eq.~\ref{eq:Dingle} to the maxima of the oscillations for D4 whereas for all other frequencies a moving window Fourier transform was used to extract a field dependent amplitude (squares and triangles). Cross-checking the two methods for the D4 orbit yielded good agreement as shown by the grey curve and open circles. For the \Jp\ orbit (triangles) different values of the mean free path were found below and above \SI{9}{\tesla} as can be seen by the two fits with distinct slope (red solid lines).}
\label{fig:Dingle}%
\end{figure}

Magnetic breakdown can be identified by a characteristic variation of the oscillation amplitude with field, the amplitude of the fundamental frequency is expected to be reduced above the breakdown field whereas the sums and differences are expected to be present above the breakdown field only. We do observe unperturbed damping of the J3 and D4 orbit as can be seen from Fig.~\ref{fig:Dingle}. The damping of the \Jp\ orbit shows some irregularities which we discuss below but the damping is inconsistent with a reduction of this frequency above the breakdown field. In summary, this suggests that magnetic interaction is the origin for the mixing of the J3 and \Jp\ frequency.

\begin{figure}%
\includegraphics[width=.8\columnwidth]{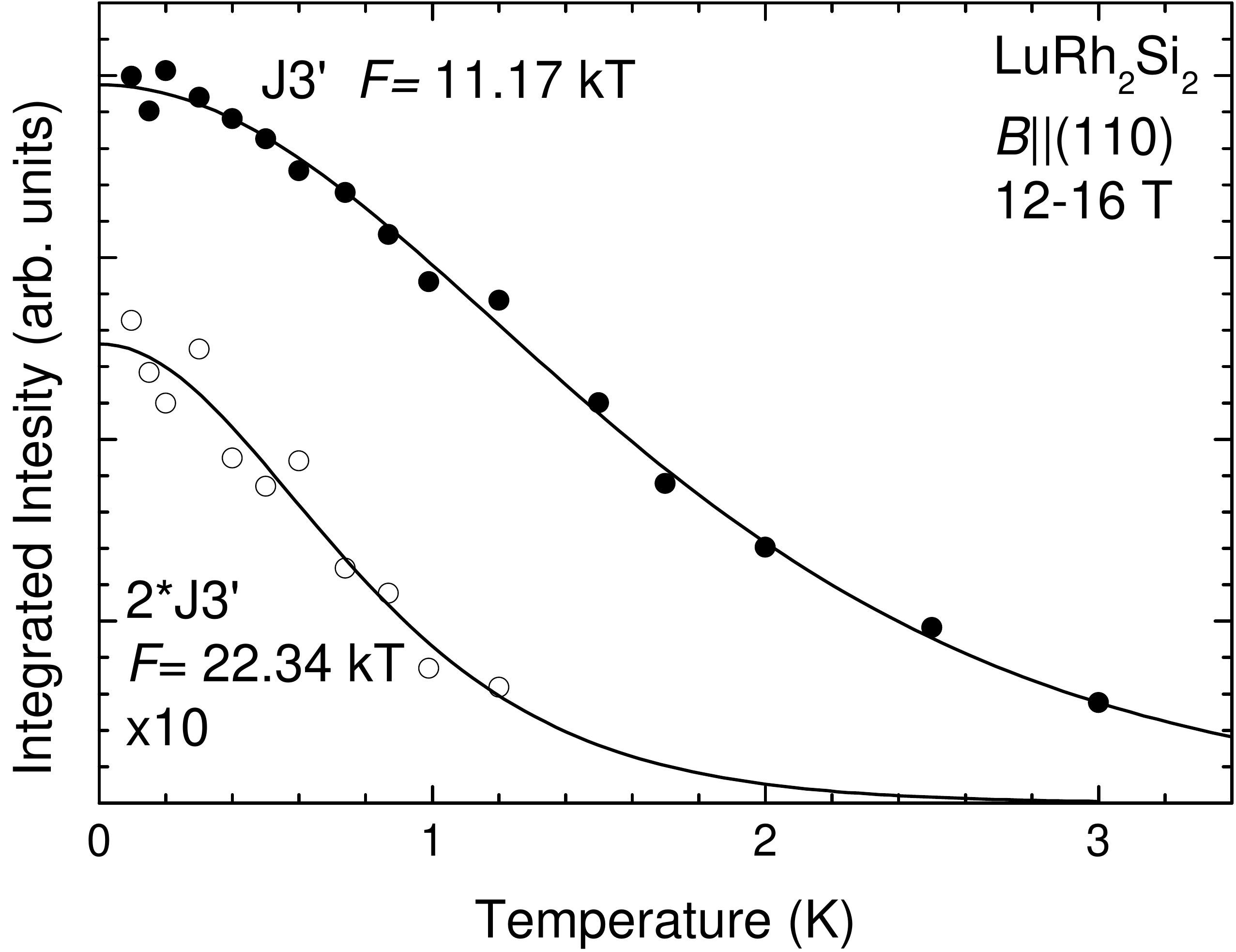}%
\caption{Lifshitz-Kosevich analysis of the temperature dependent amplitude for the  \Jp\ frequency and its harmonic $ 2*\Jp$. For the later the amplitude is enlarged by a factor of 10 for better visualisation. The effective mass was deduced by fitting Eq.~\ref{eq:mass}.
}
\label{fig:mass}%
\end{figure}

The masses for second harmonics are expected to be twice that of the fundamental and for the difference and sum the mass is expected to be the sum of the individual masses. This is not immediately consistent with the experimentally observed masses. However, we have to take into account that J3 and J8 could not be separated in the mass study. Therefore, the directly measured mass of the J3 and J8 frequency may not reflect the mass of the J3 orbit. 
Analysing the harmonics and mixed frequencies we are able to reconstruct the mass of J3:
The second harmonic of \Jp\ has twice the mass of \Jp\ (cf.\ Fig.~\ref{fig:mass} and Tab.~\ref{tab:Masses}). For the sum and difference $\Jp\pm\text J3$ we measure almost identical masses like for the second harmonic of \Jp. Consequently, this suggests that J3 has the same mass like \Jp, i.e. \SI{1.3}{$m_{\text e}$}. 
We note that the harmonic of J3 has a mass inconsistent with twice its fundamental mass which might be due to the influence of the non-separable frequencies J3 and J8 contributing in different ratios to the fundamental and harmonic frequency.

A mass of \SI{1.3}{$m_{\text e}$} for the J3 orbit is highly consistent with the mixing of this frequency with other orbits. We find mixing of J3 with D4, D3 and D2 as can be seen from the detailed analysis in Fig.~\ref{fig:Mixing} (b). In the case of the sum and difference with D4, i.e. $\text J3 \pm \text D4$ we measure similar masses of \SI{2.2}{$m_{\text e}$} and \SI{2.3}{$m_{\text e}$} which is close to the sum of the individual masses $\SI{1.3}{$m_{\text e}$} + \SI{0.8}{$m_{\text e}$}$ = \SI{2.1}{$m_{\text e}$} .
We note that the angular dependence of the frequency identified with $2*\text D4$ is also  compatible with $\Jp -D4$ (two solid black lines close by in bottom panel of Fig.~\ref{fig:Mixing} (b)); the mass off this frequency however, is twice that of D4 rather than the sum of the masses of \Jp\ and D4. Consequently, we identify this frequency with the harmonic of the D4 orbit.
This implies that the third harmonic $3*\text D4$ is very close to \Jp. However, $3*\text D4$ cannot account for the \Jp\ orbit as it is limited to a much narrower angular range and is also slightly higher in frequency (cf.\ second panel from bottom in Fig.~\ref{fig:Mixing} (b)).

We can rule out magnetic breakdown to yield the observed frequencies $\text J3 \pm \text D4$ as the orbits of J3 and D4 are well separated in $k$-space (cf.\ Fig.~\ref{fig:FS}). This is in line with the earlier conclusion that also J3 and \Jp\ mix via magnetic interaction. 

It remains unclear why  we observe two frequencies J3 and \Jp\ rather than only J3. One is certainly the J3 orbit on the J sheet. The other one, however, is not predicted by the band structure calculations. The absolute value of the predicted frequency matches best with the \Jp. However, the angular dependence favours the lower frequency with a small offset. As this is also the frequency with the higher amplitude we assign the \Freq{10.78} frequency with the J3 orbit.

We can rule out twinning and crystal domains to give rise to the two frequencies J3 and \Jp. The would not yield a merging of the two at $\phi \approx \SI{30}{\degree}$. If two domains were aligned along the $c$-direction but misaligned in the basal plane this would yield a continuously shrinking difference of the two frequencies for $\phi \to \SI{0}{\degree}$.

For magnetic breakdown one expects deviations of the field dependence of the amplitude from the Dingle behaviour described by eq. \ref{eq:Dingle}. Indeed, we find anomalies in the Dingle analysis of \Jp. This is illustrated in Fig.~\ref{fig:Dingle}. Two distinct slopes are present for \Jp\ in the logarithmic representation against $1/B$. This is in contrast to  all other frequencies that are strong enough and separable for a Dingle analysis, they show a single slope over the full range investigated.  
The change of slope for \Jp\ is showing an increased amplitude above \SI{9}{\tesla}. The small deviations of J3 from exponential damping (linear behaviour in Fig.~\ref{fig:Dingle}) around \SI{10}{\tesla} are most likely due to the presence of the secondary peak at this frequency that also shows up as a perturbation in the mass analysis. In particular, the slope at very high and small fields are identical ruling out magnetic breakdown to enhance of decrease the amplitude of this frequency due to magnetic breakdown.

The enhanced amplitude of \Jp\ above \SI{9}{\tesla} is a good indicator for magnetic breakdown to yield the \Jp\ frequency although it remains unclear which orbits are involved. The orbit J7 associated with the pillar in the J sheet is one possibility. An orbit around the pillbox of the P band might come into play when the spin majority branch starts to populate in high magnetic fields. For both cases, however, an anisotropic behaviour would be expected for rotations in the basal plane in contrast to the vanishing splitting between J3 and \Jp\ in the rotation towards (100), $\phi \to \SI{0}{\degree}$. Interestingly, if magnetic breakdown proves to be the origin of \Jp\ this allows to study the combination of magnetic breakdown yielding \Jp\ and magnetic interaction of J3 and \Jp\ in \LRS.


%
%
\section{Discussion}
\label{sec:Dis}

%
%
\subsection{Comparison to \YRS}
\label{subsec:Comp_YRS}
Our electronic structure studies on \LRS\ provide important refinement relevant for understanding \YRS. First, we find that $\zSi = \SI{0.379}{c}$ should be used for accurate band structure calculations. Here, we present refined LDA band structure calculations on \YRS. The sensitivity to \zSi\ mostly affects calculations of the ``small'' Fermi surface configuration in which the $f$ electrons were treated as core electrons not hybridizing with the conduction electrons. This configuration parallels the natural configuration of \LRS\ for which the $f$-electrons form a completely filled shell. LDA  calculations of this configuration were used in Refs.~\onlinecite{Knebel2006, Rourke2008, Sutton2010} for comparison with quantum oscillation measurements. 
The high frequency of \Freq{14} reported in Ref.~\onlinecite{Sutton2010} for fields along the (100) direction has been assigned to an orbit of the J-sheet of the ''small`` Fermi surface configuration. In general this frequency cannot be mapped to any of the orbits on the D-sheet as these are limited to below \Freq{8} in both the ''small`` and ``large'' Fermi surface calculation. Nevertheless the assignment with a branch predicted for the ``small'' Fermi surface calculation needs to be revised in the light of our results on \LRS. 

\begin{figure}%
\includegraphics[width=.8\columnwidth]{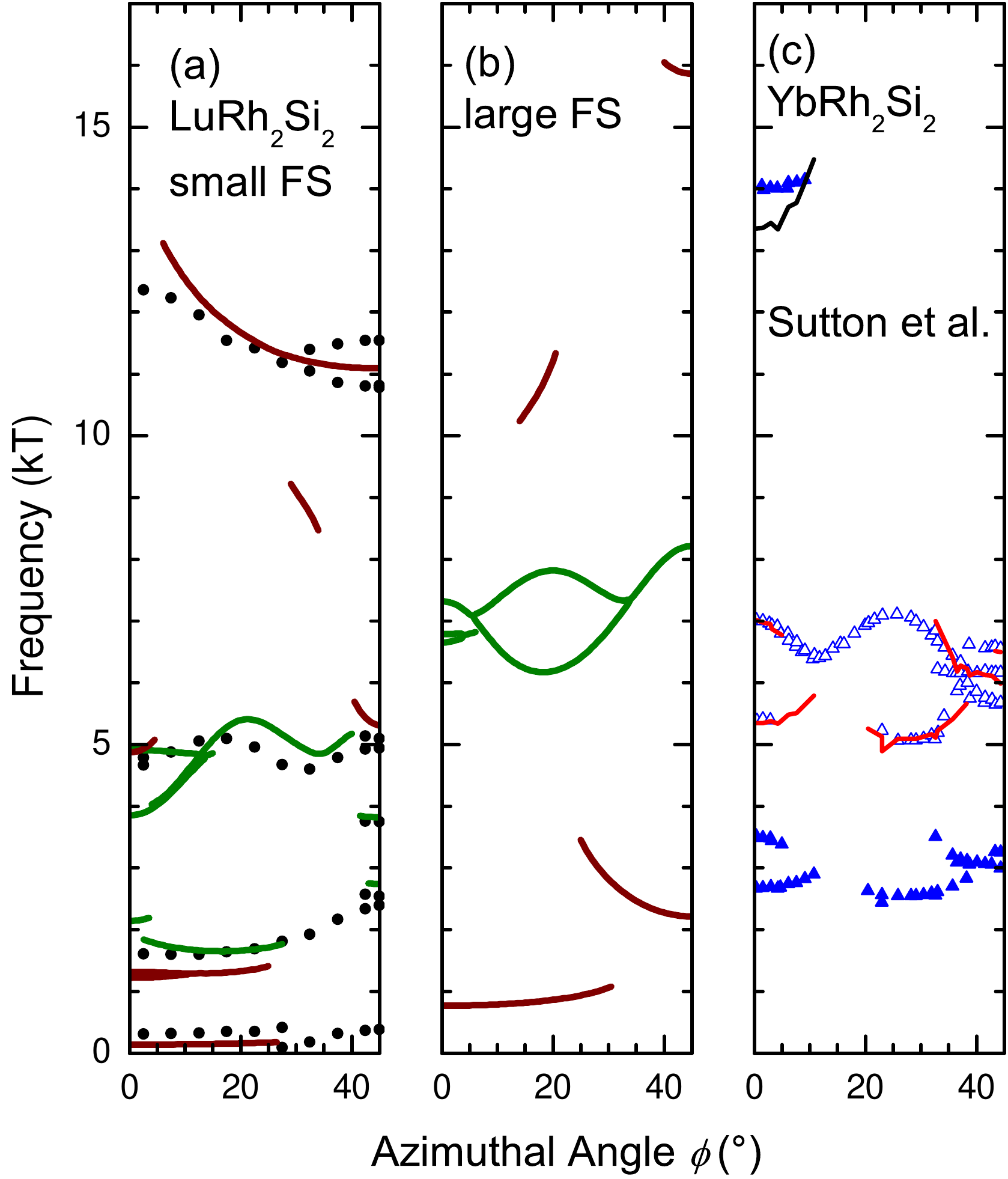}%
\caption{Comparison of quantum oscillation measurements on \LRS\ (a) and \YRS\ (c) \cite{Sutton2010} with band structure calculations of \YRS\ treating the $f$ states as core electrons (``small'' FS, (a)) and as fully itinerant (``large'' FS, (b)). Red and black lines in (c) denote second and fifth harmonics of the fundamental frequencies below \Freq{4}. Open symbols mark the frequencies possibly originating from harmonics.}
\label{fig:YRS}%
\end{figure}

In Fig.~\ref{fig:YRS} we present a comparison of refined band structure calculations on \YRS\ with the quantum oscillation measurements of Ref.~\onlinecite{Sutton2010}. We adopt identical parameters as used in Ref.~\onlinecite{Sutton2010} except for the refined lattice parameters (cf.\ Tab.~\ref{tab:latpar}). Slight variations of $c$ originate in minute variations of the Rh content \cite{Wirth2012}, we employ an average value of $c=\SI{9.86}{\angstrom}$ for our calculations on \YRS. In the case of the ``small'' Fermi surface configuration (simulated by calculating \LRS\ utilizing crystal lattice parameters of \YRS) we find virtually no difference to the results using lattice parameters of \LRS\ (cf.\ Fig.~\ref{fig:AngDep}). Consequently, the \SdH\ measurements on \LRS\ complemented by the band structure calculations can directly be used as a ``small'' Fermi surface reference of \YRS. Importantly, within this ``small'' Fermi surface calculation of \YRS\ no high-frequency orbit is predicted for field along (100). The J3 orbit is observed in the \SdH\ measurements on \LRS\ all the way to (100) but has an opposing angular dependence than the branch observed in \YRS. This indicates that the \Freq{14} frequency reported in Ref.~\onlinecite{Sutton2010} may not be assigned to the ``small'' Fermi surface configuration.

 Likewise, the ``large'' Fermi surface calculations as simulated in Refs.~\cite{Sutton2010,Rourke2008} with fully itinerant LDA calculations of \YRS\ are sensitively affected by the choice of \zSi. Using the refined lattice parameters listed in Tab.~\ref{tab:latpar} we find some major changes as can be seen from Fig.~\ref{fig:YRS}: A central hole emerges in the J sheet, which limits extremal orbits encircling the  sheet to angles between \SIlist{15;20}{\degree} -- this limits the \Freq{11} frequency to a narrow angular range (cf.\ Fig.~\ref{fig:YRS}(b)). A new extremal orbit through this central hole arises for angles above \SI{40}{\degree}, which extends beyond the boundary of the first Brillouin zone and has a high frequency of \Freq{16}.

The low frequency branches and the extremal orbits of the D-sheet are not affected by the change to the experimental lattice parameters. The orbits on the D-sheet (green lines in Fig.~\ref{fig:YRS} (b)) were earlier identified with all orbits observed in \YRS\ below \Freq{7}. 
In the light of the extensive presence of harmonics in \LRS\ we re-examine the data on \YRS. We include second and fifth harmonics of the frequencies below \Freq{4} as red and black solid lines, respectively, in Fig.~\ref{fig:YRS} (c). The frequencies between \SIlist{5;7}{\kT} very well match the angular dependence of the second harmonic. This matching extends also for rotations towards the (001) direction reported in Ref.~\onlinecite{Rourke2008}. For two of the three putative harmonics the masses reported in Ref.~\onlinecite{Sutton2010} show the expected scaling (cf.\ eq~\ref{eq:mass}), as summarized in tab.~\ref{tab:YRS_masses}.

 The fact that the higher frequencies were detected over larger angular ranges might be due to a stronger amplitude of the harmonics that may arise from magnetic interaction, as for instance is seen in CeRhIn$_5$ \cite{Schoenberg2009,Shishido2005}. We note that the modulation technique used for the de Haas-van Alphen measurements on \YRS\ in Refs.~\onlinecite{Rourke2008,Sutton2010} favours the detection of higher frequencies and harmonics ; for a typical modulation field of \SI{10}{\mT} the Bessel function determining the amplitude of the quantum oscillations in the field modulation technique yields a damping factor of $\approx 4$ for the frequencies below \Freq{4} with respect to those between \Freq{5} and \Freq{7}.

\begin{table}
  \begin{tabular}{cc c@{\hspace{1em}} cc}
    \hline
    \multicolumn{2}{c}{2nd harmonic of } && \multicolumn{2}{c}{detected frequencies} \\
        \multicolumn{2}{c}{  frequencies below \Freq{4}} && \multicolumn{2}{c}{between \Freq{5} and \Freq{7}}\\
    \cline{1-2}  \cline{4-5}
    $2*F$ & $2*\mstar / \me$ && $F$ & $\mstar / \me$\\
    \cline{1-2} \cline{4-5}
    5.3 & 23 && 5.37 & 9.2 \\
    7.0 & 12 && 7.01 & 12.3 \\
    6.4 & 14 && 6.54 & 13.2 \\
    \hline
  \end{tabular}
   \caption{Frequencies and masses expected for 2n harmonics of the frequencies below \Freq{4} compared to measured frequencies and masses between \Freq{5}  and \Freq{7} in \YRS\ after Ref.~\onlinecite{Sutton2010}.}
  \label{tab:YRS_masses}
\end{table}


In Ref.~\onlinecite{Rourke2008} it was suggested that the frequencies below and between \Freq{5} and \Freq{7} arise from orbits through the central hole of the D sheet and  slightly off-centre orbits spanning the full cross-section, respectively. In principle, this can yield roughly a factor 2 between the two groups of frequencies. However, the precise matching of the angular dependence of the putative harmonic apparent in Fig.~\ref{fig:YRS} (c) gives a very low upper boundary for the hole in the D sheet as it would otherwise violate the matching of the high frequencies with the second harmonics. In addition, the scaling should break down for rotations towards the $c$-direction. An earlier de Haas-van Alphen study cover a range up to \SI{60}{\degree} out of plane over which the frequencies seem fit with the scaling whereas the masses do deviate from the expected scaling \cite{Rourke2008}. In order to distinguish between the two possibilities we suggest quantum oscillation measurements extending the angular range all the way to $c$ axis. Here, the de Haas-van Alphen measurement of the susceptibility is unfavourable as the susceptibility is strongly reduced for this direction. This magnetic anisotropy, however, is favourable for quantum oscillation measurements using torque magnetometry. Alternatively, Shubnikov-de Haas may be used. 

If the frequencies between \Freq{5} and \Freq{7} prove to be truly harmonics this reduces the basis for comparison with band structure calculations to the observed frequencies below \Freq{4} and above \Freq{14}, which are not matched by a harmonic of the low frequencies (cf.\ black line in Fig.~\ref{fig:YRS} (c)).

Overall, the agreement of the data on \YRS\ with the LDA calculations of both the ``small'' and ``large'' Fermi surface is rather limited, particularly with the branches between \Freq{5} and \Freq{7} possibly arising from harmonics and thus not available for comparison with band structure calculations. 
The remaining fundamental frequencies below \Freq{4} and have neither a good agreement with the ``small'' nor the ``large'' Fermi surface calculation. Likewise the high frequency of $\approx \Freq{14}$, which appears to be a fundamental frequency, cannot be mapped to orbits of the LDA calculations, yet due to the fact that the largest orbits on the D-sheet is well below \Freq{9} this frequency very likely originates from the J-sheet.
LDA is very well capable to predict the Fermi surface and expected quantum oscillation frequencies of normal metals like \LRS, the ``small'' Fermi surface configuration is such a normal case. For Yb-based heavy fermion materials, however, it is known to fall short \cite{Herbst1984,Norman2005,Wigger2007}. 
Consequently, the discrepancy of our refined calculations of the ``small'' Fermi surface with earlier de Haas-van Alphen measurements on \YRS\ rules out the ``small'' Fermi surface configuration and supports the influence of the $f$ electrons in this regime. 
While the discrepancy with the LDA band structure calculations of the ``large'' Fermi surface are expected, they reinforce the need for more sophisticated models to accurately predict the electronic structure of \YRS\ and to match the observed angular dependence of quantum oscillation frequencies.

%
%
\subsection{Two-Band Character in Hall Effect of \LRS}
\label{subsec:Comp_Hall}
In Hall effect measurements on \LRS\ a pronounced crossover of the Hall coefficient as a function of temperature was found\cite{Friedemann2010c}. Through comparison with band structure calculations this could be attributed to a crossover between regimes with different relative scattering rates for the two dominating bands. The analysis in terms of a two-band model revealed similar scattering rates for the two bands at low temperatures, while they differ significantly at high temperatures. In our \SdH\ measurements we were able to extract the mean free path for the different orbits at low temperature as illustrated in Fig.~\ref{fig:Dingle}. We find the mean free path extracted for different fundamental orbits to agree within \SI{20}{\percent} only, the \Jp\ orbit shows deviations at high fields (exceeding the field scale of the Hall effect measurements). This agrees well with the result obtained through the two-band analysis.

While the two-band model gives a very convincing qualitative description of the Hall effect including the temperature range of the crossover, a small quantitative difference remains at high temperatures which could only be resolved assuming slightly different values for the Hall coefficients of the two major bands compared to the outcome of previous band structure calculations \cite{Friedemann2010c}. The previous electronic structure calculations were based on a generic $\zSi=\SI{0.375}{c}$. Our refined band structure calculations might correct for this small difference.

We note that the magnetic breakdown possibly contributing to the mixing of the various frequencies cannot account for the change in slope of the Hall resistivity \cite{Friedemann2010}. The Hall measurements were conducted with fields along the (001) direction whereas frequency mixing is seen for fields in the basal plane only. Also, the crossover in mean free path observed for the \Jp\ orbit is observed for fields along the (110) direction and at much higher fields than the crossover in Hall effect \cite{Friedemann2010c,Friedemann2010}.

Likewise it is unlikely, that a thermally excited population of the P sheet at temperatures of the order of \SI{100}{\kelvin} accounts for the change in Hall coefficient as a function of temperature. The charge carrier concentration of this band will be very small compared to the other bands and its effect is therefore negligible.

%
%
\section{Conclusion}
\label{sec:Conclusion}
We present a comprehensive study of the electronic structure of \LRS\ which---owing to almost identical lattice parameters---serves as an ideal non-magnetic reference for the intensively studied heavy-fermion material \YRS. We find a sensitive dependence of the Fermi surface topology on the position of the Si atoms \zSi.  Best agreement between predicted and measured quantum oscillation frequencies is obtained at the experimental value $\zSi=\SI{0.379}{c}$ very close to the value \SI{0.381}{c} obtained from lattice relaxation. We  therefore recommend usage of the precisely determined experimental lattice parameters for future band structure calculations on \LRS\ and \YRS. 

The re-examination of  de Haas-van Alphen measurements on \YRS\ suggests previously unidentified harmonics which reduce the number of  fundamental frequencies to a group of frequencies below \Freq{4} and a single frequency at \Freq{14}. We compare these frequencies with both the results on \LRS\ and LDA calculations, which are well capable to describe the ''small`` Fermi surface configuration within  the $f$-core treatment. 
This comparison reveals strong deviation which support the earlier conclusion that the \YRS\ $f$ electrons do not localize at $\mu_0 H_0 \approx \SI{10}{\tesla}$.
%
%
\begin{acknowledgments}
The authors would like to thank Gil Lonzarich and Stephen Julian for fruitful discussions. We thank Yuri Grin for crystal structure refinement and Lina Klintberg and Cornelius Krellner for technical support. SF acknowledges support by the Alexander von Humboldt foundation and ERC, PR acknowledges support by the Cusanuswerk and the EPSRC. his work was partially supported by NSF-DMR-0710492, NSF-PHY-0551164, FP7-ERC-227378.
\end{acknowledgments}


\begin{thebibliography}{31}
\expandafter\ifx\csname natexlab\endcsname\relax\def\natexlab#1{#1}\fi
\expandafter\ifx\csname bibnamefont\endcsname\relax
  \def\bibnamefont#1{#1}\fi
\expandafter\ifx\csname bibfnamefont\endcsname\relax
  \def\bibfnamefont#1{#1}\fi
\expandafter\ifx\csname citenamefont\endcsname\relax
  \def\citenamefont#1{#1}\fi
\expandafter\ifx\csname url\endcsname\relax
  \def\url#1{\texttt{#1}}\fi
\expandafter\ifx\csname urlprefix\endcsname\relax\def\urlprefix{URL }\fi
\providecommand{\bibinfo}[2]{#2}
\providecommand{\eprint}[2][]{\url{#2}}

\bibitem[{\citenamefont{Gegenwart et~al.}(2002)\citenamefont{Gegenwart,
  Custers, Geibel, Neumaier, Tayama, Tenya, Trovarelli, and
  Steglich}}]{Gegenwart2002}
\bibinfo{author}{\bibfnamefont{P.}~\bibnamefont{Gegenwart}},
  \bibinfo{author}{\bibfnamefont{J.}~\bibnamefont{Custers}},
  \bibinfo{author}{\bibfnamefont{C.}~\bibnamefont{Geibel}},
  \bibinfo{author}{\bibfnamefont{K.}~\bibnamefont{Neumaier}},
  \bibinfo{author}{\bibfnamefont{T.}~\bibnamefont{Tayama}},
  \bibinfo{author}{\bibfnamefont{K.}~\bibnamefont{Tenya}},
  \bibinfo{author}{\bibfnamefont{O.}~\bibnamefont{Trovarelli}},
  \bibnamefont{and} \bibinfo{author}{\bibfnamefont{F.}~\bibnamefont{Steglich}},
  \bibinfo{journal}{Phys. Rev. Lett.} \textbf{\bibinfo{volume}{89}},
  \bibinfo{pages}{56402} (\bibinfo{year}{2002}).

\bibitem[{\citenamefont{Si et~al.}(2001)\citenamefont{Si, Rabello, Ingersent,
  and Smith}}]{Si2001}
\bibinfo{author}{\bibfnamefont{Q.}~\bibnamefont{Si}},
  \bibinfo{author}{\bibfnamefont{M.~S.} \bibnamefont{Rabello}},
  \bibinfo{author}{\bibfnamefont{K.}~\bibnamefont{Ingersent}},
  \bibnamefont{and} \bibinfo{author}{\bibfnamefont{J.~L.} \bibnamefont{Smith}},
  \bibinfo{journal}{Nature} \textbf{\bibinfo{volume}{413}},
  \bibinfo{pages}{804} (\bibinfo{year}{2001}).

\bibitem[{\citenamefont{Paschen et~al.}(2004)\citenamefont{Paschen,
  L\"{u}hmann, Wirth, Gegenwart, Trovarelli, Geibel, Steglich, Coleman, and
  Si}}]{Paschen2004}
\bibinfo{author}{\bibfnamefont{S.}~\bibnamefont{Paschen}},
  \bibinfo{author}{\bibfnamefont{T.}~\bibnamefont{L\"{u}hmann}},
  \bibinfo{author}{\bibfnamefont{S.}~\bibnamefont{Wirth}},
  \bibinfo{author}{\bibfnamefont{P.}~\bibnamefont{Gegenwart}},
  \bibinfo{author}{\bibfnamefont{O.}~\bibnamefont{Trovarelli}},
  \bibinfo{author}{\bibfnamefont{C.}~\bibnamefont{Geibel}},
  \bibinfo{author}{\bibfnamefont{F.}~\bibnamefont{Steglich}},
  \bibinfo{author}{\bibfnamefont{P.}~\bibnamefont{Coleman}}, \bibnamefont{and}
  \bibinfo{author}{\bibfnamefont{Q.}~\bibnamefont{Si}},
  \bibinfo{journal}{Nature} \textbf{\bibinfo{volume}{432}},
  \bibinfo{pages}{881} (\bibinfo{year}{2004}).

\bibitem[{\citenamefont{Friedemann
  et~al.}(2010{\natexlab{a}})\citenamefont{Friedemann, Oeschler, Wirth,
  Krellner, Geibel, Steglich, Paschen, Kirchner, and Si}}]{Friedemann2010b}
\bibinfo{author}{\bibfnamefont{S.}~\bibnamefont{Friedemann}},
  \bibinfo{author}{\bibfnamefont{N.}~\bibnamefont{Oeschler}},
  \bibinfo{author}{\bibfnamefont{S.}~\bibnamefont{Wirth}},
  \bibinfo{author}{\bibfnamefont{C.}~\bibnamefont{Krellner}},
  \bibinfo{author}{\bibfnamefont{C.}~\bibnamefont{Geibel}},
  \bibinfo{author}{\bibfnamefont{F.}~\bibnamefont{Steglich}},
  \bibinfo{author}{\bibfnamefont{S.}~\bibnamefont{Paschen}},
  \bibinfo{author}{\bibfnamefont{S.}~\bibnamefont{Kirchner}}, \bibnamefont{and}
  \bibinfo{author}{\bibfnamefont{Q.}~\bibnamefont{Si}}, \bibinfo{journal}{Proc.
  Natl. Acad. Sci.} \textbf{\bibinfo{volume}{107}}, \bibinfo{pages}{14547}
  (\bibinfo{year}{2010}{\natexlab{a}}).

\bibitem[{\citenamefont{Friedemann
  et~al.}(2010{\natexlab{b}})\citenamefont{Friedemann, Wirth, Oeschler,
  Krellner, Geibel, Steglich, MaQuilon, Fisk, Paschen, and
  Zwicknagl}}]{Friedemann2010c}
\bibinfo{author}{\bibfnamefont{S.}~\bibnamefont{Friedemann}},
  \bibinfo{author}{\bibfnamefont{S.}~\bibnamefont{Wirth}},
  \bibinfo{author}{\bibfnamefont{N.}~\bibnamefont{Oeschler}},
  \bibinfo{author}{\bibfnamefont{C.}~\bibnamefont{Krellner}},
  \bibinfo{author}{\bibfnamefont{C.}~\bibnamefont{Geibel}},
  \bibinfo{author}{\bibfnamefont{F.}~\bibnamefont{Steglich}},
  \bibinfo{author}{\bibfnamefont{S.}~\bibnamefont{MaQuilon}},
  \bibinfo{author}{\bibfnamefont{Z.}~\bibnamefont{Fisk}},
  \bibinfo{author}{\bibfnamefont{S.}~\bibnamefont{Paschen}}, \bibnamefont{and}
  \bibinfo{author}{\bibfnamefont{G.}~\bibnamefont{Zwicknagl}},
  \bibinfo{journal}{Phys. Rev. B} \textbf{\bibinfo{volume}{82}},
  \bibinfo{pages}{35103} (\bibinfo{year}{2010}{\natexlab{b}}).

\bibitem[{\citenamefont{Caroca-Canales}(2010)}]{Caroca-Canales2010}
\bibinfo{author}{\bibfnamefont{N.}~\bibnamefont{Caroca-Canales}},
  \bibinfo{journal}{private communication}  (\bibinfo{year}{2010}).

\bibitem[{\citenamefont{Cardoso}(2011)}]{Cardoso2011}
\bibinfo{author}{\bibfnamefont{R.}~\bibnamefont{Cardoso}},
  \bibinfo{journal}{private communication}  (\bibinfo{year}{2011}).

\bibitem[{\citenamefont{Wirth et~al.}(2012)\citenamefont{Wirth, Ernst,
  Cardoso-Gil, Borrmann, Seiro, Krellner, Geibel, Kirchner, Burkhardt, Grin
  et~al.}}]{Wirth2012}
\bibinfo{author}{\bibfnamefont{S.}~\bibnamefont{Wirth}},
  \bibinfo{author}{\bibfnamefont{S.}~\bibnamefont{Ernst}},
  \bibinfo{author}{\bibfnamefont{R.}~\bibnamefont{Cardoso-Gil}},
  \bibinfo{author}{\bibfnamefont{H.}~\bibnamefont{Borrmann}},
  \bibinfo{author}{\bibfnamefont{S.}~\bibnamefont{Seiro}},
  \bibinfo{author}{\bibfnamefont{C.}~\bibnamefont{Krellner}},
  \bibinfo{author}{\bibfnamefont{C.}~\bibnamefont{Geibel}},
  \bibinfo{author}{\bibfnamefont{S.}~\bibnamefont{Kirchner}},
  \bibinfo{author}{\bibfnamefont{U.}~\bibnamefont{Burkhardt}},
  \bibinfo{author}{\bibfnamefont{Y.}~\bibnamefont{Grin}}, \bibnamefont{et~al.},
  \bibinfo{journal}{Journal of physics: Condensed matter}
  \textbf{\bibinfo{volume}{24}}, \bibinfo{pages}{294203}
  (\bibinfo{year}{2012}).

\bibitem[{\citenamefont{Pfau et~al.}(2013)\citenamefont{Pfau, Daou, Lausberg,
  Naren, Brando, Friedemann, Wirth, Westerkamp, Stockert, Gegenwart
  et~al.}}]{Pfau2013}
\bibinfo{author}{\bibfnamefont{H.}~\bibnamefont{Pfau}},
  \bibinfo{author}{\bibfnamefont{R.}~\bibnamefont{Daou}},
  \bibinfo{author}{\bibfnamefont{S.}~\bibnamefont{Lausberg}},
  \bibinfo{author}{\bibfnamefont{H.~R.} \bibnamefont{Naren}},
  \bibinfo{author}{\bibfnamefont{M.}~\bibnamefont{Brando}},
  \bibinfo{author}{\bibfnamefont{S.}~\bibnamefont{Friedemann}},
  \bibinfo{author}{\bibfnamefont{S.}~\bibnamefont{Wirth}},
  \bibinfo{author}{\bibfnamefont{T.}~\bibnamefont{Westerkamp}},
  \bibinfo{author}{\bibfnamefont{U.}~\bibnamefont{Stockert}},
  \bibinfo{author}{\bibfnamefont{P.}~\bibnamefont{Gegenwart}},
  \bibnamefont{et~al.}, \bibinfo{journal}{arXiv:1302.6867 [cond-mat.str-el]}
  p.~\bibinfo{pages}{5} (\bibinfo{year}{2013}), \eprint{1302.6867}.

\bibitem[{\citenamefont{Rourke et~al.}(2008)\citenamefont{Rourke, McCollam,
  Lapertot, Knebel, Flouquet, and Julian}}]{Rourke2008}
\bibinfo{author}{\bibfnamefont{P.~M.~C.} \bibnamefont{Rourke}},
  \bibinfo{author}{\bibfnamefont{A.}~\bibnamefont{McCollam}},
  \bibinfo{author}{\bibfnamefont{G.}~\bibnamefont{Lapertot}},
  \bibinfo{author}{\bibfnamefont{G.}~\bibnamefont{Knebel}},
  \bibinfo{author}{\bibfnamefont{J.}~\bibnamefont{Flouquet}}, \bibnamefont{and}
  \bibinfo{author}{\bibfnamefont{S.~R.} \bibnamefont{Julian}},
  \bibinfo{journal}{Phys. Rev. Lett.} \textbf{\bibinfo{volume}{101}},
  \bibinfo{pages}{237205} (\bibinfo{year}{2008}).

\bibitem[{\citenamefont{Gegenwart et~al.}(2006)\citenamefont{Gegenwart, Tokiwa,
  Westerkamp, Weickert, Custers, Ferstl, Krellner, Geibel, Kerschl, M\"{u}ller
  et~al.}}]{Gegenwart2006}
\bibinfo{author}{\bibfnamefont{P.}~\bibnamefont{Gegenwart}},
  \bibinfo{author}{\bibfnamefont{Y.}~\bibnamefont{Tokiwa}},
  \bibinfo{author}{\bibfnamefont{T.}~\bibnamefont{Westerkamp}},
  \bibinfo{author}{\bibfnamefont{F.}~\bibnamefont{Weickert}},
  \bibinfo{author}{\bibfnamefont{J.}~\bibnamefont{Custers}},
  \bibinfo{author}{\bibfnamefont{J.}~\bibnamefont{Ferstl}},
  \bibinfo{author}{\bibfnamefont{C.}~\bibnamefont{Krellner}},
  \bibinfo{author}{\bibfnamefont{C.}~\bibnamefont{Geibel}},
  \bibinfo{author}{\bibfnamefont{P.}~\bibnamefont{Kerschl}},
  \bibinfo{author}{\bibfnamefont{K.~H.} \bibnamefont{M\"{u}ller}},
  \bibnamefont{et~al.}, \bibinfo{journal}{New J. Phys.}
  \textbf{\bibinfo{volume}{8}}, \bibinfo{pages}{171} (\bibinfo{year}{2006}).

\bibitem[{\citenamefont{Zwicknagl}(2011)}]{Zwicknagl2011}
\bibinfo{author}{\bibfnamefont{G.}~\bibnamefont{Zwicknagl}},
  \bibinfo{journal}{Journal of Physics: Condensed Matter}
  \textbf{\bibinfo{volume}{23}}, \bibinfo{pages}{94215} (\bibinfo{year}{2011}).

\bibitem[{\citenamefont{Sutton et~al.}(2010)\citenamefont{Sutton, Rourke,
  Taufour, McCollam, Lapertot, Knebel, Flouquet, and Julian}}]{Sutton2010}
\bibinfo{author}{\bibfnamefont{A.~B.} \bibnamefont{Sutton}},
  \bibinfo{author}{\bibfnamefont{P.~M.~C.} \bibnamefont{Rourke}},
  \bibinfo{author}{\bibfnamefont{V.}~\bibnamefont{Taufour}},
  \bibinfo{author}{\bibfnamefont{A.}~\bibnamefont{McCollam}},
  \bibinfo{author}{\bibfnamefont{G.}~\bibnamefont{Lapertot}},
  \bibinfo{author}{\bibfnamefont{G.}~\bibnamefont{Knebel}},
  \bibinfo{author}{\bibfnamefont{J.}~\bibnamefont{Flouquet}}, \bibnamefont{and}
  \bibinfo{author}{\bibfnamefont{S.~R.} \bibnamefont{Julian}},
  \bibinfo{journal}{Phys. Status Solidi B} \textbf{\bibinfo{volume}{247}},
  \bibinfo{pages}{549} (\bibinfo{year}{2010}).

\bibitem[{\citenamefont{Herbst and Wilkins}(1984)}]{Herbst1984}
\bibinfo{author}{\bibfnamefont{J.}~\bibnamefont{Herbst}} \bibnamefont{and}
  \bibinfo{author}{\bibfnamefont{J.}~\bibnamefont{Wilkins}},
  \bibinfo{journal}{Physical Review B} \textbf{\bibinfo{volume}{29}},
  \bibinfo{pages}{5992} (\bibinfo{year}{1984}).

\bibitem[{\citenamefont{Knebel et~al.}(2006)\citenamefont{Knebel, Boursier,
  Hassinger, Lapertot, Niklowitz, Pourret, Salce, Sanchez, Sheikin, Bonville
  et~al.}}]{Knebel2006}
\bibinfo{author}{\bibfnamefont{G.}~\bibnamefont{Knebel}},
  \bibinfo{author}{\bibfnamefont{R.}~\bibnamefont{Boursier}},
  \bibinfo{author}{\bibfnamefont{E.}~\bibnamefont{Hassinger}},
  \bibinfo{author}{\bibfnamefont{G.}~\bibnamefont{Lapertot}},
  \bibinfo{author}{\bibfnamefont{P.~G.} \bibnamefont{Niklowitz}},
  \bibinfo{author}{\bibfnamefont{A.}~\bibnamefont{Pourret}},
  \bibinfo{author}{\bibfnamefont{B.}~\bibnamefont{Salce}},
  \bibinfo{author}{\bibfnamefont{J.~P.} \bibnamefont{Sanchez}},
  \bibinfo{author}{\bibfnamefont{I.}~\bibnamefont{Sheikin}},
  \bibinfo{author}{\bibfnamefont{P.}~\bibnamefont{Bonville}},
  \bibnamefont{et~al.}, \bibinfo{journal}{J. Phys. Soc. Jpn.}
  \textbf{\bibinfo{volume}{75}}, \bibinfo{pages}{114709}
  (\bibinfo{year}{2006}).

\bibitem[{\citenamefont{Suzuki and Harima}(2010)}]{Suzuki2010}
\bibinfo{author}{\bibfnamefont{M.-T.} \bibnamefont{Suzuki}} \bibnamefont{and}
  \bibinfo{author}{\bibfnamefont{H.}~\bibnamefont{Harima}},
  \bibinfo{journal}{Journal of the Physical Society of Japan}
  \textbf{\bibinfo{volume}{79}}, \bibinfo{pages}{024705}
  (\bibinfo{year}{2010}).

\bibitem[{\citenamefont{Kimber et~al.}(2009)\citenamefont{Kimber, Kreyssig,
  Zhang, Jeschke, Valent\'{\i}, Yokaichiya, Colombier, Yan, Hansen, Chatterji
  et~al.}}]{Kimber2009}
\bibinfo{author}{\bibfnamefont{S.~A.~J.} \bibnamefont{Kimber}},
  \bibinfo{author}{\bibfnamefont{A.}~\bibnamefont{Kreyssig}},
  \bibinfo{author}{\bibfnamefont{Y.-Z.} \bibnamefont{Zhang}},
  \bibinfo{author}{\bibfnamefont{H.~O.} \bibnamefont{Jeschke}},
  \bibinfo{author}{\bibfnamefont{R.}~\bibnamefont{Valent\'{\i}}},
  \bibinfo{author}{\bibfnamefont{F.}~\bibnamefont{Yokaichiya}},
  \bibinfo{author}{\bibfnamefont{E.}~\bibnamefont{Colombier}},
  \bibinfo{author}{\bibfnamefont{J.}~\bibnamefont{Yan}},
  \bibinfo{author}{\bibfnamefont{T.~C.} \bibnamefont{Hansen}},
  \bibinfo{author}{\bibfnamefont{T.}~\bibnamefont{Chatterji}},
  \bibnamefont{et~al.}, \bibinfo{journal}{Nature materials}
  \textbf{\bibinfo{volume}{8}}, \bibinfo{pages}{471} (\bibinfo{year}{2009}).

\bibitem[{\citenamefont{Lee et~al.}(2008)\citenamefont{Lee, Iyo, Eisaki, Kito,
  Fernandez-Diaz, Ito, Kihou, Matsuhata, Braden, and Yamada}}]{Lee2008}
\bibinfo{author}{\bibfnamefont{C.-H.} \bibnamefont{Lee}},
  \bibinfo{author}{\bibfnamefont{A.}~\bibnamefont{Iyo}},
  \bibinfo{author}{\bibfnamefont{H.}~\bibnamefont{Eisaki}},
  \bibinfo{author}{\bibfnamefont{H.}~\bibnamefont{Kito}},
  \bibinfo{author}{\bibfnamefont{M.~T.} \bibnamefont{Fernandez-Diaz}},
  \bibinfo{author}{\bibfnamefont{T.}~\bibnamefont{Ito}},
  \bibinfo{author}{\bibfnamefont{K.}~\bibnamefont{Kihou}},
  \bibinfo{author}{\bibfnamefont{H.}~\bibnamefont{Matsuhata}},
  \bibinfo{author}{\bibfnamefont{M.}~\bibnamefont{Braden}}, \bibnamefont{and}
  \bibinfo{author}{\bibfnamefont{K.}~\bibnamefont{Yamada}},
  \bibinfo{journal}{Journal of the Physical Society of Japan}
  \textbf{\bibinfo{volume}{77}}, \bibinfo{pages}{083704}
  (\bibinfo{year}{2008}).

\bibitem[{\citenamefont{Wigger et~al.}(2007)\citenamefont{Wigger, Baumberger,
  Shen, Yin, Pickett, Maquilon, and Fisk}}]{Wigger2007}
\bibinfo{author}{\bibfnamefont{G.~A.} \bibnamefont{Wigger}},
  \bibinfo{author}{\bibfnamefont{F.}~\bibnamefont{Baumberger}},
  \bibinfo{author}{\bibfnamefont{Z.~X.} \bibnamefont{Shen}},
  \bibinfo{author}{\bibfnamefont{Z.~P.} \bibnamefont{Yin}},
  \bibinfo{author}{\bibfnamefont{W.~E.} \bibnamefont{Pickett}},
  \bibinfo{author}{\bibfnamefont{S.}~\bibnamefont{Maquilon}}, \bibnamefont{and}
  \bibinfo{author}{\bibfnamefont{Z.}~\bibnamefont{Fisk}},
  \bibinfo{journal}{Phys. Rev. B} \textbf{\bibinfo{volume}{76}},
  \bibinfo{pages}{35106} (\bibinfo{year}{2007}).

\bibitem[{\citenamefont{Jeong}(2006)}]{Jeong2006}
\bibinfo{author}{\bibfnamefont{T.}~\bibnamefont{Jeong}}, \bibinfo{journal}{J.
  Phys.: Condens. Matter} \textbf{\bibinfo{volume}{18}}, \bibinfo{pages}{10529}
  (\bibinfo{year}{2006}).

\bibitem[{\citenamefont{Reiss et~al.}(2013)\citenamefont{Reiss, Rourke,
  Zwicknagl, Grosche, and Friedemann}}]{Reiss2013}
\bibinfo{author}{\bibfnamefont{P.}~\bibnamefont{Reiss}},
  \bibinfo{author}{\bibfnamefont{P.~M.~C.} \bibnamefont{Rourke}},
  \bibinfo{author}{\bibfnamefont{G.}~\bibnamefont{Zwicknagl}},
  \bibinfo{author}{\bibfnamefont{F.~M.} \bibnamefont{Grosche}},
  \bibnamefont{and}
  \bibinfo{author}{\bibfnamefont{S.}~\bibnamefont{Friedemann}},
  \bibinfo{journal}{physica status solidi (b)} \textbf{\bibinfo{volume}{250}},
  \bibinfo{pages}{498} (\bibinfo{year}{2013}).

\bibitem[{\citenamefont{Blaha et~al.}(2011)\citenamefont{Blaha, Schwarz,
  Madsen, Kvasnicka, and Luitz}}]{Blaha2011}
\bibinfo{author}{\bibfnamefont{P.}~\bibnamefont{Blaha}},
  \bibinfo{author}{\bibfnamefont{K.}~\bibnamefont{Schwarz}},
  \bibinfo{author}{\bibfnamefont{G.~K.~H.} \bibnamefont{Madsen}},
  \bibinfo{author}{\bibfnamefont{D.}~\bibnamefont{Kvasnicka}},
  \bibnamefont{and} \bibinfo{author}{\bibfnamefont{J.}~\bibnamefont{Luitz}},
  \emph{\bibinfo{title}{{An Augmented Plane Wave + Local Orbitals Program for
  Calculating Crystal Properties}}} (\bibinfo{year}{2011}).

\bibitem[{\citenamefont{Perdew et~al.}(1996)\citenamefont{Perdew, Burke, and
  Ernzerhof}}]{Perdew1996}
\bibinfo{author}{\bibfnamefont{J.~P.} \bibnamefont{Perdew}},
  \bibinfo{author}{\bibfnamefont{K.}~\bibnamefont{Burke}}, \bibnamefont{and}
  \bibinfo{author}{\bibfnamefont{M.}~\bibnamefont{Ernzerhof}},
  \bibinfo{journal}{Physical Review Letters} \textbf{\bibinfo{volume}{77}},
  \bibinfo{pages}{3865} (\bibinfo{year}{1996}).

\bibitem[{\citenamefont{Kokalj}(1999)}]{Kokalj1999}
\bibinfo{author}{\bibfnamefont{A.}~\bibnamefont{Kokalj}},
  \bibinfo{journal}{Journal of Molecular Graphics and Modelling}
  \textbf{\bibinfo{volume}{17}}, \bibinfo{pages}{176} (\bibinfo{year}{1999}).

\bibitem[{\citenamefont{Rourke and Julian}(2012)}]{Rourke2012}
\bibinfo{author}{\bibfnamefont{P.~M.~C.} \bibnamefont{Rourke}}
  \bibnamefont{and} \bibinfo{author}{\bibfnamefont{S.~R.}
  \bibnamefont{Julian}}, \bibinfo{journal}{Computer Physics Communications}
  \textbf{\bibinfo{volume}{183}}, \bibinfo{pages}{324} (\bibinfo{year}{2012}).

\bibitem[{\citenamefont{Rourke et~al.}(2009)\citenamefont{Rourke, McCollam,
  Lapertot, Knebel, Flouquet, and Julian}}]{Rourke2009}
\bibinfo{author}{\bibfnamefont{P.~M.~C.} \bibnamefont{Rourke}},
  \bibinfo{author}{\bibfnamefont{A.}~\bibnamefont{McCollam}},
  \bibinfo{author}{\bibfnamefont{G.}~\bibnamefont{Lapertot}},
  \bibinfo{author}{\bibfnamefont{G.}~\bibnamefont{Knebel}},
  \bibinfo{author}{\bibfnamefont{J.}~\bibnamefont{Flouquet}}, \bibnamefont{and}
  \bibinfo{author}{\bibfnamefont{S.~R.} \bibnamefont{Julian}},
  \bibinfo{journal}{Journal of Physics: Conference Series}
  \textbf{\bibinfo{volume}{150}}, \bibinfo{pages}{042165}
  (\bibinfo{year}{2009}).

\bibitem[{\citenamefont{Maquilon}(2007)}]{Maquilon2007}
\bibinfo{author}{\bibfnamefont{S.}~\bibnamefont{Maquilon}}, Ph.D. thesis,
  \bibinfo{school}{UC Davis} (\bibinfo{year}{2007}).

\bibitem[{\citenamefont{Shoenberg}(2009)}]{Schoenberg2009}
\bibinfo{author}{\bibfnamefont{D.}~\bibnamefont{Shoenberg}},
  \emph{\bibinfo{title}{{Magnetic Oscillations in Metals}}}
  (\bibinfo{publisher}{Cambridge University Press},
  \bibinfo{address}{Cambridge}, \bibinfo{year}{2009}).

\bibitem[{\citenamefont{Shishido et~al.}(2005)\citenamefont{Shishido, Settai,
  Harima, and Onuki}}]{Shishido2005}
\bibinfo{author}{\bibfnamefont{H.}~\bibnamefont{Shishido}},
  \bibinfo{author}{\bibfnamefont{R.}~\bibnamefont{Settai}},
  \bibinfo{author}{\bibfnamefont{H.}~\bibnamefont{Harima}}, \bibnamefont{and}
  \bibinfo{author}{\bibfnamefont{Y.}~\bibnamefont{Onuki}}, \bibinfo{journal}{J.
  Phys. Soc. Jpn.} \textbf{\bibinfo{volume}{74}}, \bibinfo{pages}{1103}
  (\bibinfo{year}{2005}).

\bibitem[{\citenamefont{Norman}(2005)}]{Norman2005}
\bibinfo{author}{\bibfnamefont{M.~R.~G.} \bibnamefont{Norman}},
  \bibinfo{journal}{Phys. Rev. B} \textbf{\bibinfo{volume}{71}},
  \bibinfo{pages}{220405(R)} (\bibinfo{year}{2005}).

\bibitem[{\citenamefont{Friedemann
  et~al.}(2010{\natexlab{c}})\citenamefont{Friedemann, Oeschler, Wirth,
  Steglich, MaQuilon, and Fisk}}]{Friedemann2010}
\bibinfo{author}{\bibfnamefont{S.}~\bibnamefont{Friedemann}},
  \bibinfo{author}{\bibfnamefont{N.}~\bibnamefont{Oeschler}},
  \bibinfo{author}{\bibfnamefont{S.}~\bibnamefont{Wirth}},
  \bibinfo{author}{\bibfnamefont{F.}~\bibnamefont{Steglich}},
  \bibinfo{author}{\bibfnamefont{S.}~\bibnamefont{MaQuilon}}, \bibnamefont{and}
  \bibinfo{author}{\bibfnamefont{Z.}~\bibnamefont{Fisk}},
  \bibinfo{journal}{Phys. Status Solidi B} \textbf{\bibinfo{volume}{247}},
  \bibinfo{pages}{723} (\bibinfo{year}{2010}{\natexlab{c}}).

\end{thebibliography}

\end{document}